\documentclass[%
 reprint,
superscriptaddress,
 amsmath,amssymb,
 aps,
]{revtex4-2}

\usepackage{graphicx}
\usepackage{dcolumn}
\usepackage{bm}
\usepackage{hyperref}

\usepackage{dsfont}
\usepackage{braket}
\usepackage{xcolor}

\begin{document}

\preprint{APS/123-QED}

\title{Trion-phonon interaction in atomically thin semiconductors}

\author{Raul Perea-Causin}
\email{causin@chalmers.se}
\affiliation{Department of Physics, Chalmers University of Technology, 412 96 Gothenburg, Sweden}
\author{Samuel Brem}
\affiliation{Department of Physics,
Philipps-Universität Marburg, 35032 Marburg, Germany}
\author{Ermin Malic}
\affiliation{Department of Physics,
Philipps-Universität Marburg, 35032 Marburg, Germany}
\affiliation{Department of Physics, Chalmers University of Technology, 412 96 Gothenburg, Sweden}

\begin{abstract}
Optical and transport properties of doped monolayer semiconductors are dominated by trions, which are three-particle compounds formed by two electrons and one hole or vice versa.  
In this work, we investigate the trion-phonon interaction on a microscopic footing and apply our model to the exemplary case of a molybdenum diselenide (MoSe\textsubscript{2}) monolayer.
We determine the trion series of states and their internal quantum structure by solving the trion Schrödinger equation.
Transforming the system into a trion basis and solving equations of motion, including the trion-phonon interaction within the second-order Born-Markov approximation, provides a microscopic access to the trion dynamics.
In particular, we investigate trion propagation and compute the diffusion coefficient and mobility.
In the low density limit, we find that trions propagate less efficiently than excitons and electrons due to their stronger coupling with phonons and their larger mass.
For increasing densities, we predict a drastic enhancement of diffusion caused by the build-up of a large pressure by the degenerate trion gas, which is a direct consequence of the fermionic character of trions. Our work provides microscopic insights into the trion-phonon interaction and its impact on trion transport in atomically thin semiconductors.
\end{abstract}

\maketitle

\section{Introduction}
Atomically thin semiconductors, with the prominent example of transition metal dichalcogenide (TMD) monolayers, have emerged in the last years both as a platform for investigating fundamental many-particle quantum phenomena as well as a promising candidate for novel optoelectronic applications\,\cite{yu2015valley,wang2018colloquium,mueller2018exciton,jin2018ultrafast,tran2019evidence}.
The strong Coulomb interaction in these materials favours the formation of excitons---tightly-bound electron-hole pairs---, which dominate  optics, dynamics, and transport properties in undoped TMDs\,\cite{wang2018colloquium,mueller2018exciton}.
In most materials, however, doping appears either unintentionally due to impurities\,\cite{docherty2014ultrafast} or intentionally via e.g. a gate voltage\,\cite{mak2013tightly,ross2013electrical}.
In the p (n) doping regime, the photoexcited electron-hole pairs bind to doping charges and form positive (negative) trions---three-particles complexes consisting of two electrons (holes) and one hole (electron).
Thus, in presence of doping, trions are expected to govern optical and transport properties of TMDs.

So far, experimental and theoretical studies of doped TMDs have focused mostly on investigating optical properties. It has been found that the optical absorption is highly tunable with doping and is dominated by Fermi-polarons\,\cite{sidler2017fermi,efimkin2017many,glazov2020optical,rana2020many,imamoglu2021exciton,efimkin2021electron,katsch2022excitonic}.
Moreover, the rich landscape of bright, dark and excited trion states has been observed in optical absorption and emission spectra\,\cite{plechinger2016trion,courtade2017charged,arora2019excited,arora2020dark,wagner2020autoionization,goldstein2020ground,he2020valley,liu2021exciton,yang2022relaxation,klein2022trions} and partially understood with different theoretical approaches\,\cite{deilmann2017dark,fey2020theory,katsch2022excitonic,yang2022relaxation,klein2022trions}.
Other studies have explored the dynamics of trion formation\,\cite{singh2016trion}, recombination\,\cite{wang2016radiative}, and valley depolarization\,\cite{plechinger2016trion,singh2016long}. More recently, trion propagation has been also investigated, reporting a relatively fast\,\cite{uddin2020neutral,kim2021free} and long\,\cite{kato2016transport,cadiz2018exciton} diffusion, as well as thermal\,\cite{park2021imaging} and electric\,\cite{cheng2021observation} drift.

\begin{figure}[t!]
    \centering
    \includegraphics[width=\linewidth]{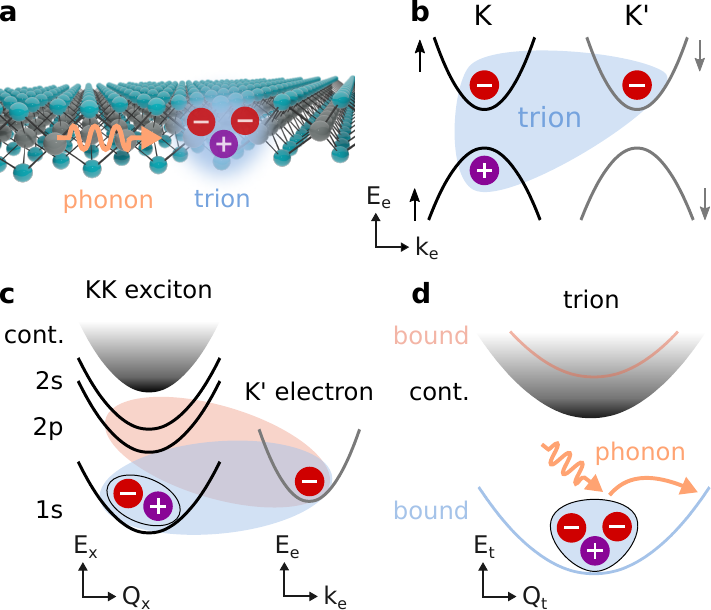}
    \caption{\textbf{a} Illustration of trion-phonon interaction in a TMD monolayer.
     Trion configuration in the \textbf{b} electron-hole  and  
    \textbf{c} exciton-electron picture for a MoSe\textsubscript{2} monolayer. The blue (red) shaded area illustrates the mostly 1s (2s/2p) exciton character of the ground (excited) trion state. Note that the energy axes $E_\text{x}$ and $E_\text{e}$ are different and the relative position of electron and exciton bands cannot be compared.
    \textbf{d} Trion eigenstates consisting of the ground (blue) and excited (red) bound states as well as the scattering continuum. A trion can scatter within its center-of-mass dispersion by absorbing or emitting a phonon (cf. orange arrow).}
    \label{fig1}
\end{figure}

While trion-phonon scattering is expected to be important for charge transport in TMDs at low and even moderate trion densities, only recently the interaction between trions and phonons has been approached\,\cite{ayari2020phonon, zipfel2022electron}. In particular, trion cooling in MoSe\textsubscript{2} monolayers due to scattering with acoustic phonons has been reported\,\cite{zipfel2022electron}, providing a quantitative estimate of the trion-phonon scattering rate in these materials.
The characterization of trion-phonon scattering is essential for understanding trion thermalization, cooling, and transport, which play a crucial role in potential applications exploiting the large oscillator strength of excitons and the non-zero charge of trions.
Despite the importance of trion-phonon interaction, little is known about the microscopic nature of such a complex process where a three-particle compound interacts with a phonon.

In this work, we investigate the trion-phonon interaction and its impact on trion propagation in TMD monolayers (cf. Fig.\,\ref{fig1}a), with particular focus on the influence of the trion substructure and its fermionic nature.
Solving the trion Schrödinger equation, we obtain a microscopic access to the trion eigenstates and their internal quantum structure.
With the calculated trion wave functions we evaluate the trion-phonon coupling strength and investigate trion diffusion.
In comparison with excitons and single electrons, we find that trion diffusion at low densities is rather slow because of the strong coupling with phonons and the large trion mass.
Interestingly, we predict a considerable enhancement of the trion diffusion at low temperatures and high densities due to the build-up of a large pressure gradient by the degenerate trion gas.

\section{Microscopic model}
We consider a molybdenum diselenide (MoSe\textsubscript{2}) monolayer, which exhibits maxima and minima of the spin-split valence and conduction bands at the K and K' high-symmetry points of the Brillouin zone (cf. Fig.\ref{fig1}b).
Due to the small electron-hole mass imbalance in this material, trions with positive and negative charge will share similar properties. Therefore, it is sufficient to study only one trion species.
In this work, we consider n-doped TMD samples, where trions are formed by two electrons and one hole, cf. Fig.\,\ref{fig1}b.

We develop a microscopic model to describe trion dynamics and, in particular, address the trion-phonon interaction.
The hallmark of our model is the trion Hamiltonian, which allows us to describe the dynamics of the trion occupation by exploting Heisenberg's equation of motion. While the derivation of the trion Hamiltonian is thoroughly presented in Appendix\,\ref{SI:Ht}, here we summarize the main steps.
The starting point is the Hamilton operator of the electron-hole system depicted in Fig.\,\ref{fig1}b,
\begin{align}
    H =& \sum_{\mathbf{k}} \left( E^{\text{e}}_{\mathbf{k}} e^{\dagger}_{\mathbf{k}} e^{\phantom{\dagger}}_{\mathbf{k}} + E^{\text{e'}}_{\mathbf{k}} e'^{\dagger}_{\mathbf{k}} e'^{\phantom{\dagger}}_{\mathbf{k}} +  E^{\text{h}}_{\mathbf{k}} h^{\dagger}_{\mathbf{k}} h^{\phantom{\dagger}}_{\mathbf{k}}\right) \nonumber \\
    & + \sum_{\mathbf{kk}\mathbf{'q}} V_{\mathbf{q}} \left( e^{\dagger}_{\mathbf{k}+\mathbf{q}} e'^{\dagger}_{\mathbf{k}'-\mathbf{q}} e'^{\phantom{\dagger}}_{\mathbf{k}'} e^{\phantom{\dagger}}_{\mathbf{k}} - e^{\dagger}_{\mathbf{k}+\mathbf{q}} h^{\dagger}_{\mathbf{k}'-\mathbf{q}} h^{\phantom{\dagger}}_{\mathbf{k}'} e^{\phantom{\dagger}}_{\mathbf{k}} \right. \nonumber \\
    & \left. - e'^{\dagger}_{\mathbf{k}+\mathbf{q}} h^{\dagger}_{\mathbf{k}'-\mathbf{q}} h^{\phantom{\dagger}}_{\mathbf{k}'} e'^{\phantom{\dagger}}_{\mathbf{k}} \right),
    \label{eq:Hel}
\end{align}
including the Coulomb interaction $V_\mathbf{q}$ between charges.
Here, the operators $e_\mathbf{k}^{({}_{'})}$ and $h_\mathbf{k}$ annihilate a K\textsuperscript{(}'\textsuperscript{)} electron and a K hole with the momentum $\mathbf{k}$ and single-particle energies $E^{\text{e\textsuperscript{(}'\textsuperscript{)}}}_{\mathbf{k}}$ and $E^{\text{h}}_{\mathbf{k}}$, respectively.
We expand the electron/hole creation and annihilation operators in the Fock subspace of single trions---in analogy to the method used in Refs.\,\cite{katsch2018theory,ivanov1993self} to describe excitons.
Next, we transform the system into the exciton-electron picture depicted in Fig.\,\ref{fig1}c by expanding electron-hole creation/annihilation operators into an exciton basis.
This allows us to resolve the excitonic character of trion states. Finally, we expand the appearing exciton-electron creation/annihilation operators into a trion basis and obtain the Hamilton operator for a free trion,
\begin{equation}
    H_{\text{t,0}} = \sum_{\lambda\mathbf{Q}_\text{t}} \left( \varepsilon_\text{t}^{\lambda} + \frac{\hbar^2\mathbf{Q}_\text{t}^2}{2M_\text{t}} \right) T^{\dagger}_{\lambda,\mathbf{Q}_\text{t}} T^{\phantom{\dagger}}_{\lambda,\mathbf{Q}_\text{t}}.
    \label{eq:Ht}
\end{equation}
This Hamiltonian describes trions with the energy $\varepsilon_\text{t}^{\lambda} + \hbar^2\mathbf{Q}_\text{t}^2/(2M_\text{t})$ (cf. Fig\,\ref{fig1}d), where $M_\text{t}=M_\text{x}+m_\text{e}$ and $M_\text{x}=m_\text{e}+m_\text{h}$ are the trion and exciton masses, respectively. The electron/hole effective masses, $m_\text{e}=0.5m_0,m_\text{h}=0.6m_0$, are taken from ab-initio calculations \,\cite{kormanyos2015k}.
In Eq.\,\eqref{eq:Ht} we have introduced the trion annihilation (creation) operators $T^{(\dagger)}_{\lambda,\mathbf{Q}_\text{t}}$ which destroy (create) a trion at the eigenstate $\lambda$ with the center-of-mass momentum $\mathbf{Q}_\text{t}$.
These operators are defined as
$T^{(\dagger)}_{\lambda,\mathbf{Q}_\text{t}}=\sum_{\nu\mathbf{k}}\psi^{\lambda *}_{\nu\mathbf{k}} X^{(\dagger)}_{\nu\mathbf{Q}_x}e'^{(\dagger)}_{\mathbf{k}_e}$, where $\psi^{\lambda}_{\nu,\mathbf{k}}$ is the wave function of the trion state $\lambda$ with the relative exciton-electron momentum $\mathbf{k}$.
The different excitonic quantum numbers contributing to the trion are denoted with the index $\nu$.
The operator $X^{(\dagger)}_{\nu\mathbf{Q}_\text{x}}e'^{(\dagger)}_{\mathbf{k}_\text{e}}$ destroys (creates) an exciton at state $\nu$ with the momentum $\mathbf{Q}_\text{x}=\beta_\text{x}\mathbf{Q}_\text{t}-\mathbf{k}$ and an electron at the K' valley with the momentum $\mathbf{k}_\text{e}=\beta_\text{e}\mathbf{Q}_\text{t}-\mathbf{k}$, where $\beta_\text{e}=m_\text{e}/M_\text{t}$ and $\beta_\text{x}=M_\text{x}/M_\text{t}$.
The exchange interaction splitting singlet and triplet states\,\cite{yu2014dirac,courtade2017charged,klein2022trions} is beyond the scope of this work but can be principally incorporated into our theory.
Note also that the trion picture considered here is equivalent to the Fermi-polaron picture at sufficiently low densities\,\cite{glazov2020optical,zipfel2022electron}.

The trion eigenenergies $\varepsilon^\lambda_\text{t}$ and wave functions $\psi^{\lambda}_{\nu,\mathbf{k}}$ are obtained by solving the three-body Schrödinger equation in exciton-electron basis,
\begin{equation}
    \left( \varepsilon_\text{x}^{\nu}+\frac{\hbar^2\mathbf{k}^2}{2m_\text{x-e}} \right) \psi^{\lambda}_{\nu,\mathbf{k}} + \sum_{\mu\mathbf{q}} \tilde{V}^{\nu\mu}_{\mathbf{q}} \psi^{\lambda}_{\mu,\mathbf{k}+\mathbf{q}} = \varepsilon_\text{t}^{\lambda} \psi^{\lambda}_{\nu,\mathbf{k}}.
\label{eq:trionSchrodinger}
\end{equation}
This equation describes the eigenstates formed by an exciton-electron compound and is completely analogous to the Schrödinger equation for two electrons and one hole.
The first term in Eq.\,\eqref{eq:trionSchrodinger} accounts for the relative motion of
the Coulomb-bound electron-hole pair with the exciton binding energy $\varepsilon_\text{x}^{\nu}$,
along with the relative motion of 
a non-interacting exciton-electron compound with the relative mass $m_\text{x-e}=M_\text{x}m_\text{e}/M_\text{t}$.
The second term describes the Coulomb interaction between the exciton and the electron with the matrix element
$\tilde{V}^{\nu\mu}_{\mathbf{q}} = V_{\mathbf{q}}\braket{\nu|\left(\mathrm{e}^{i\alpha_\text{h}\mathbf{q \cdot r}}-\mathrm{e}^{-i\alpha_\text{e}\mathbf{q \cdot r}}\right)|\mu}$ and leads to a mixing of exciton states, i.e. a polarization of the exciton within the trion. Here $\alpha_{e,h}=m_{e,h}/M_\text{x}$ has been introduced and $V_\mathbf{q}$ corresponds to the Coulomb potential in the TMD for which we have adopted the model from Ref. \onlinecite{van2018coulomb} that accurately reproduces the screening dependence of the trion binding energies\,.

\begin{figure}[t!]
    \centering
    \includegraphics[width=\linewidth]{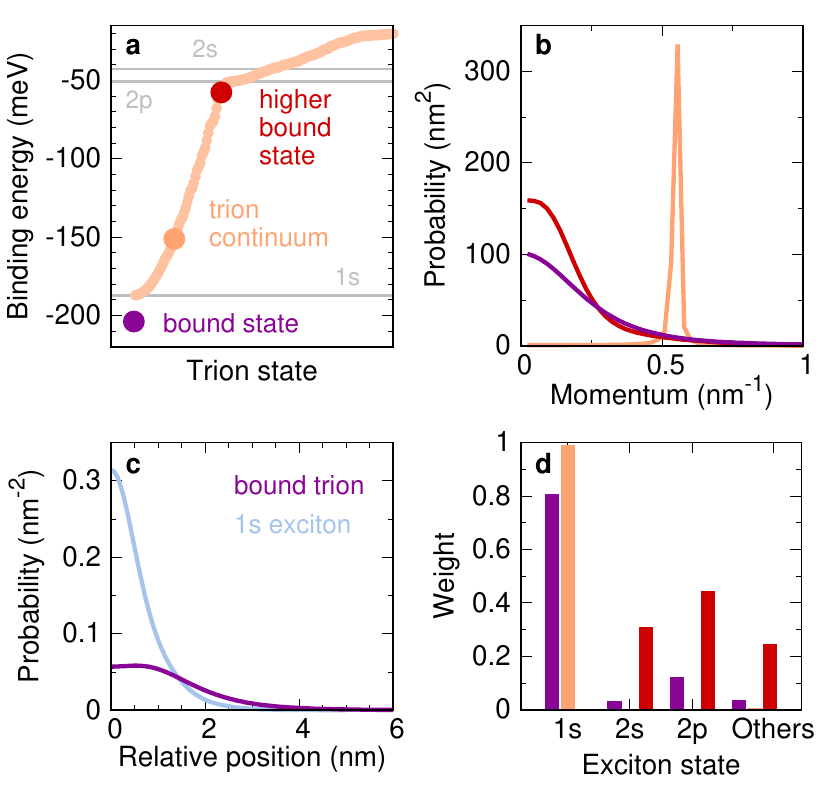}
    \caption{Trion eigenstates in hBN-encapsulated MoSe\textsubscript{2} monolayer.
    \textbf{a} Trion eigenenergies vs. eigenstate number $\lambda$, with the ground (higher) bound state marked by a purple (red) dot and the trion continuum states denoted in orange. The 1s, 2p, and 2s exciton binding energies are shown as a reference.
    \textbf{b} Trion probability distribution $P^{\lambda}_{\mathbf{k}}$ as a function of the relative exciton-electron momentum $\mathbf{k}$ for the ground and the higher bound trion state, as well as for an exemplary continuum state denoted with an orange dot in part a.
    \textbf{c} Probability distribution for the trion (exciton) ground state as a function of the exciton-electron (electron-hole) separation.
    \textbf{d} Excitonic weights $p^\lambda_\nu$ (i.e. probability that the exciton within the trion state $\lambda$ is occupying the state $\nu$) for the three considered trion states.}
    \label{fig2}
\end{figure}

The main goal of our study is to model the trion-phonon interaction. To this end, we transform the electron-phonon and exciton-phonon interaction into a trion basis and obtain the trion-phonon Hamiltonian (cf. Appendix\,\ref{SI:Htp} for more details),
\begin{equation}
    H_\text{t-p} = \sum_{\mathbf{Q}_\text{t}\mathbf{q}\lambda\lambda'} G^{\lambda\lambda'}_{\text{t},\mathbf{q}} T^{\dagger}_{\lambda,\mathbf{Q}_\text{t}+\mathbf{q}} T^{\phantom{\dagger}}_{\lambda',\mathbf{Q}_\text{t}} \left( b^{\phantom{\dagger}}_{\mathbf{q}} + b^{\dagger}_{-\mathbf{q}} \right),
    \label{eq:Htp}
\end{equation}
with phonon annihilation (creation) operators, $b^{(\dagger)}_{\mathbf{q}}$, and the trion-phonon matrix element,
\begin{equation}
    G^{\lambda\lambda'}_{\text{t},\mathbf{q}} = \sum_{\mathbf{k}\nu\mu} \psi^{\lambda'*}_{\mu,\mathbf{k}} \left(  G^{\nu\mu}_{\text{x},\mathbf{q}} \psi^\lambda_{\nu,\mathbf{k}+\beta_\text{e}\mathbf{q}} + \delta_{\nu\mu}g^\text{e}_\mathbf{q} \psi^\lambda_{\nu,\mathbf{k}-\beta_\text{x}\mathbf{q}} \right),
    \label{eq:trion-phonon}
\end{equation}
which describes a trion transition from $\lambda'$ to $\lambda$ with the momentum transfer $\mathbf{q}$ due to the emission or absorption of a phonon.
We have introduced the exciton-phonon and electron-phonon matrix elements, $G^{\nu\mu}_{\text{x},\mathbf{q}}$ and $g^\text{e}_\mathbf{q}$, respectively. The electron-phonon coupling is treated in a deformation-potential approach, where the potentials for electrons and holes are assumed to be similar and are fitted to match the experimentally measured exciton linewidth in Ref.\,\cite{hotta2020exciton} (see details in Appendix\,\ref{SI:Htp}).
The first and second terms in Eq.\,\eqref{eq:trion-phonon} correspond to the exciton-phonon and electron-phonon coupling strengths, respectively, summed over all possible transitions with the momentum transfer $\mathbf{q}$ and weighted by the wave function overlap between initial and final trion states.
We note already here that we focus on the low temperature regime where intravalley scattering via long-wavelength acoustic phonons is the only relevant channel. The extension of Eq.\,\eqref{eq:Htp} to several phonon modes and multiple valleys is straightforward.

\section{Trion eigenstates}
In order to investigate trion-phonon interaction, we first need to determine the trion eigenstates and their internal quantum structure.
Thus, we solve the trion Schrödinger equation for the negatively-charged trion (cf. Fig.\,\ref{fig1}b) and plot the computed series of trion eigenstates in Fig.\,\ref{fig2}a.
Details for the treatment and numerical solution of Eq.\,\eqref{eq:trionSchrodinger} can be found in the Appendix\,\ref{SI:1D}.
We find that the ground state lies 17 meV below the 1s exciton energy (purple dot in Fig.\,\ref{fig2}a), which is in good agreement with previous theoretical studies\,\cite{van2018coulomb,katsch2022excitonic} but lower than the experimentally reported values between 25 and 30 meV\,\cite{sidler2017fermi,florian2018dielectric,zipfel2022electron}. This discrepancy has been suggested to originate from a vacuum gap between the TMD and the surrounding materials\,\cite{florian2018dielectric} or from the polaronic enhancement of the effective masses\,\cite{van2018coulomb}.
Above the 1s exciton energy, we find a continuum of trion states displaying a quadratic relation with the trion quantum number $\lambda$, which is a characteristic feature of scattering states\,\cite{fey2020theory}. While most of these states describe exciton-electron scattering, we identify one state with a clear bound nature located 16 (8) meV below the 2s (2p) exciton state (red dot in Fig.\,\ref{fig2}a). This higher bound state has been experimentally observed in optical spectra\,\cite{arora2019excited,goldstein2020ground,wagner2020autoionization,liu2021exciton}, and has also been theoretically predicted \,\cite{fey2020theory,katsch2022excitonic}.

To better understand the nature of trion states, we exploit the calculated trion wave functions. First, we compute the probability distribution for the exciton-electron relative momentum,
$ P^{\lambda}_{\mathbf{k}} = \sum_{\nu}|\psi^\lambda_{\nu,\mathbf{k}}|^2$, which is shown in Fig.\,\ref{fig2}b for the ground and higher bound states, as well as for an exemplary continuum state denoted with an orange dot in Fig.\,\ref{fig2}a.
While the bound state probabilities are centered at $\mathbf{k}=0$, the probability for the continuum state is peaked at a non-zero value corresponding to the momentum at which the exciton and the electron move apart from each other.
These characteristic features reflect the bound and scattering nature of the respective states.
The higher bound state is more confined than the ground state in momentum space, cf. Fig.\,\ref{fig2}b.  This means that the exciton-electron separation is larger, i.e. the higher state is less bound.
Furthermore, the ground trion state is more extended than the 1s exciton in real space, cf. Fig.\,\ref{fig2}c.
Concretely, the ground state trion probability distribution exhibits an average exciton-electron distance of $\sqrt{\braket{\mathbf{r}^2}} = 2.5\ \text{nm}$, while the 1s exciton displays an average electron-hole separation of 1.3 nm.

We further analyze the substructure of the trion states by evaluating the excitonic weights, $p^\lambda_\nu = \sum_{\mathbf{k}}|\psi^\lambda_{\nu,\mathbf{k}}|^2$, which describe the probability that the exciton (within the trion state $\lambda$) is in the  state $\nu$.
The excitonic weights of 1s, 2s, 2p and all the other exciton states are shown in Fig.\,\ref{fig2}d for the three trion states that have been discussed above.
We find that the ground trion state (purple bars) has mostly 1s exciton character (as illustrated in Fig.\,\ref{fig1}c), with a small but significant 2p contribution. This mixing of 1s and 2p states describes the polarization of the exciton due to the Coulomb potential of the additional electron. The attraction between the resulting dipole and the electron is the main contribution to the trion binding energy. Nevertheless, we note that considering only 1s and 2p states is far from sufficient for obtaining a converged trion binding energy.

The exemplary trion continuum state (orange bars) displays a full 1s exciton character. Thus, these states describe a 1s exciton that scatters with an electron. Although not shown here, the trion continuum states above the 2p and 2s exciton energies equivalently display a 2s and 2p character, as discussed in Ref.\,\cite{fey2020theory}.
The higher bound trion state (red bars) has been denoted in literature as 2s trion because it appears in optical spectra close to the 2s exciton resonance\,\cite{arora2019excited,goldstein2020ground,wagner2020autoionization,katsch2022excitonic}, and 2p trion because its eigenenergy lies right below the 2p exciton resonance\,\cite{fey2020theory}.
Here, we find that this state is dominated by the 2p and 2s exciton states (as illustrated in Fig.\,\ref{fig1}c), with a sizable contribution of other states. The similar 2p and 2s character of this state results from the strong s-p coupling and the energetic proximity of the 2p and 2s exciton resonances.

\section{Trion-phonon interaction}
We now investigate the trion-phonon interaction and how it is influenced by the trion substructure. 
In the following, we focus on the ground trion state, considering only intra-state transitions mediated by long-wavelength acoustic phonons since these are the most efficient processes at low temperatures where trions are stable.
At higher temperatures, optical phonons become relevant and can mediate the dissociation of bound trions into the exciton-electron continuum.

\begin{figure}[t!]
    \centering
    \includegraphics[width=\linewidth]{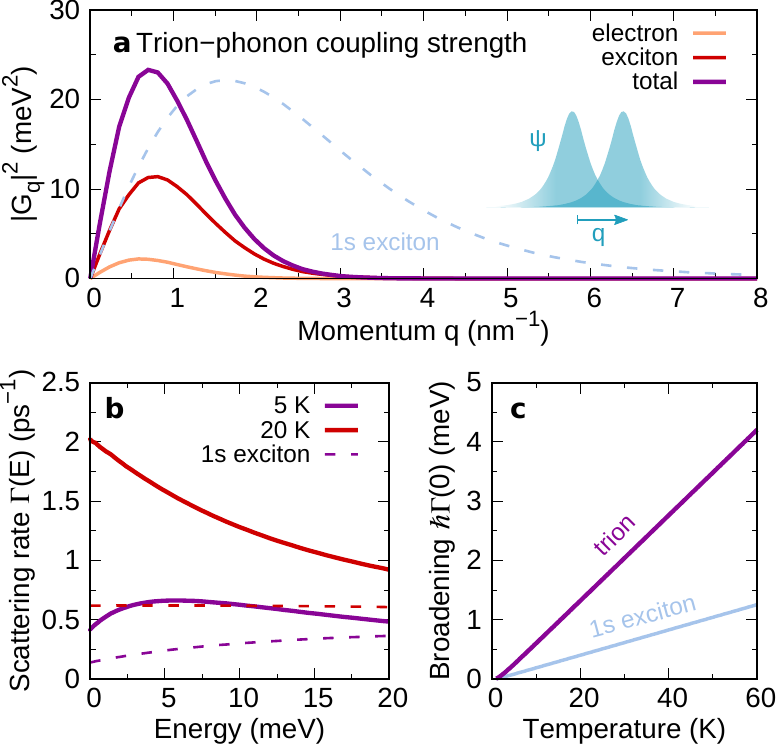}
    \caption{Trion-phonon interaction.
    \textbf{a} Trion-phonon coupling strength as a function of momentum transfer $\mathbf{q}$ explicitly showing the single electron and exciton contributions. The 1s exciton-phonon coupling strength is also shown for comparison (dashed blue-line). The inset illustrates the wave function overlap between initial and final states.
    \textbf{b} Trion-phonon scattering rates as a function of kinetic energy at 5 K  and 20 K. The corresponding rates for 1s exciton-phonon scattering are plotted with dashed lines.
    \textbf{c} Temperature-dependent phonon-induced spectral broadening of the trion and 1s exciton states.}
    \label{fig3}
\end{figure}

In Fig.\,\ref{fig3}a we show the absolute squared value of the trion-phonon matrix element, Eq.\,\eqref{eq:trion-phonon}, as a function of momentum transfer. We evaluate separately the electron-phonon and exciton-phonon terms in Eq.\,\eqref{eq:trion-phonon} (orange and red lines, respectively). The slope at small $\mathbf{q}$ directly reflects the coupling with phonons and is given by the deformation potential of each quasi-particle, which is smaller for electrons than for excitons (as for the latter it is given by the sum of the electron and hole contributions).
At higher momenta, the wave function overlap between initial and final trion states,
$\sum_{\mathbf{k}} \psi^{\lambda*}_{\mu,\mathbf{k}} \psi^\lambda_{\nu,\mathbf{k}+\mathbf{q}}$,
becomes significantly reduced (cf. the inset) and, in consequence, transitions involving a large momentum transfer are inhibited. Note that the coupling strength for the 1s exciton-phonon interaction (dashed blue line) extends to larger momenta, reflecting the longer range of the 1s exciton wave function in momentum space. The purple line in Fig. \,\ref{fig2}a shows the total trion-phonon coupling strength, which is slightly more pronounced and decays much faster than the 1s exciton-phonon coupling. Moreover, we find that the trion-phonon interaction is dominated by transitions involving 1s initial and final exciton states ($\nu,\mu=\text{1s}$), reflecting the 1s character of the exciton within the ground trion state (cf. Fig.\,\ref{fig2}d).

Next, we use the calculated coupling strength to evaluate the trion-phonon scattering rates, which are crucial for modeling and understanding trion thermalization, propagation, and cooling, among other processes. The trion-phonon Hamiltonian in Eq.\,\eqref{eq:Htp} is analogous to an electron-phonon Hamiltonian. Moreover, the trion operators obey fermionic commutation relations at sufficiently low densities (see Appendix\,\ref{SI:Ht}). Therefore, we can derive an equation of motion for the trion occupation and find the trion-phonon scattering rates in second-order Born-Markov approximation analogously to the case of electrons using the density matrix formalism\,\cite{kuhn1992monte,brem20}. Assuming low trion densities, the expression for the (out-) scattering rate reads
\begin{equation}
    \Gamma_{\mathbf{Q}_\text{t}} = \frac{2\pi}{\hbar} \sum_{\mathbf{q}\pm} \left| G_{\text{t},\mathbf{q}} \right|^2 \eta^{\pm}_{\mathbf{q}}\; \delta\!\left(E_{\mathbf{Q}_\text{t}+\mathbf{q}}-E_{\mathbf{Q}_\text{t}}\pm\hbar\Omega_{\mathbf{q}}\right)
\end{equation}
where $\eta^\pm_{\mathbf{q}}=n_{\mathbf{q}}+\frac{1}{2}\pm\frac{1}{2}$ with $n_\mathbf{q}$ being the phonon number (Bose-Einstein distribution), $E_{\mathbf{Q}_\text{t}}=\hbar^2\mathbf{Q}_\text{t}^2/(2M_\text{t})$ is the trion kinetic energy, and $\Omega_\mathbf{q}=s|\mathbf{q}|$ is the acoustic phonon frequency with the sound speed $s$ taken from Ref.\,\cite{jin2014intrinsic}. Note that we restrict the initial and final trion state $\lambda=\lambda'$ to the ground state.

In Fig.\,\ref{fig3}b we show the trion-phonon scattering rates as a function of trion kinetic energy at two different temperatures. 
At 5K the phonon number $n_\mathbf{q}$ is small and, in consequence, scattering is dominated by spontaneous emission. Phonon emission is only allowed above a threshold for the trion kinetic energy at which energy and momentum conservation of the scattering process is fulfilled. Below this threshold (i.e. at the bottom of the centre-of-mass dispersion), scattering is dominated by phonon absorption and is thus suppressed due to the small number of phonons (cf. dip in the solid purple line).
On the other hand, at 20 K the phonon number is sufficiently large so that phonon absorption and stimulated emission dominate.
In principle, these two processes contribute equally to the scattering rate, resulting in a rate that is independent on the kinetic energy of the initial trion state.
Nevertheless, scattering at high kinetic energies involves transitions with a large momentum transfer, which are inhibited due to the reduced wave function overlap between initial and final states (cf. Fig.\,\ref{fig3}a).
This manifests as a weakening of the scattering rate at higher kinetic energies (cf. red line in Fig.\,\ref{fig3}b).
This effect is much less pronounced for excitons (cf. dashed lines), which exhibit a flatter scattering rate reflecting the larger momentum extension of the exciton-phonon coupling.
Moreover, exciton-phonon scattering is significantly weaker than trion-phonon scattering due to the smaller exciton mass and the smaller coupling elements with long-wavelength acoustic phonons.
We note here that a larger trion binding energy (closer to the experimental values) would result in a larger momentum extension of the trion-phonon coupling and thus in a flatter trion-phonon scattering rate.

The scattering rate $\Gamma_{\mathbf{Q}}$ determines the energetic broadening of the state with momentum $\mathbf{Q}$. In particular, the exciton-phonon scattering rate at the light cone ($\mathbf{Q}_\text{x}\approx 0$) is responsible for the broadening of exciton resonances in optical spectra\,\cite{selig16, brem19}. 
While the broadening of the trion peak in optical spectra additionally involves the electron recoil effect\,\cite{zipfel2022electron}, we here evaluate the impact of phonons.
In Fig.\,\ref{fig3}c we plot the phonon-induced broadening of the state with zero center-of-mass momentum for trions and excitons as a function of temperature. As the trion-phonon scattering is stronger, it results in a significantly larger broadening, reaching 3.5 meV at 50 K compared to 1 meV for excitons. The linear dependence on temperature is characteristic of scattering with long-wavelength acoustic phonons\,\cite{selig16}.
Trion-phonon scattering has been reported to be responsible for the cooling of trions\,\cite{zipfel2022electron}. Based on the models from Refs.\,\cite{zipfel2022electron,kaasbjerg2014hot} but including the trion form factor (wave function overlap), we find trion cooling times in the range of 4-10 ps for temperatures up to 50 K (see Appendix\,\ref{SI:cooling}). The good agreement with experimental measurements in Ref.\,\cite{zipfel2022electron} supports the predictive character of our microscopic theory.
Another work has investigated the cooling of free electrons in MoSe\textsubscript{2} and obtained a cooling time of 70 ps\,\cite{venanzi2021terahertz}. The larger cooling time of free electrons compared to trions is consistent with the smaller electron mass and weaker coupling with phonons.

\section{Trion diffusion and mobility}
\begin{figure}[t!]
    \centering
    \includegraphics[width=0.7\linewidth]{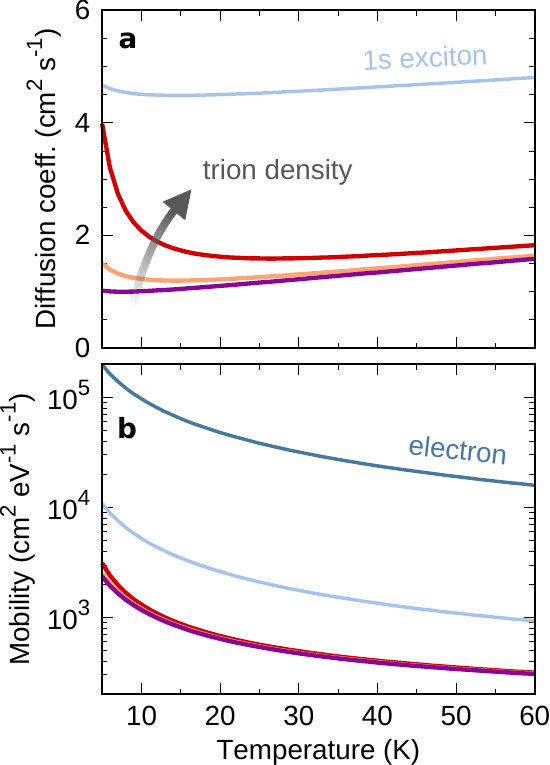}
    \caption{Trion diffusion and mobility.
    \textbf{a} Temperature-dependent trion diffusion coefficient in the low-density limit (purple) and for trion densities of $1 \times 10^{11}\ \text{cm}^{-2}$ (orange) and $4 \times 10^{11}\ \text{cm}^{-2}$ (red). The low-density 1s exciton diffusion coefficient is shown in light-blue.
    \textbf{b} Trion, 1s exciton, and electron mobilities as a function of temperature.}
    \label{fig4}
\end{figure}

Now, we investigate the impact of the internal quantum structure and fermionic character of trions on trion diffusion and mobility. The diffusion coefficient can be derived from our microscopic approach in a relaxation time approximation\,\cite{hess1996maxwell} and reads
\begin{equation}
    D = -\frac{1}{2A}\sum_{\mathbf{Q}} \tau_{\mathbf{Q}} \mathbf{v}_{\mathbf{Q}}^2 \frac{\partial\rho_{\mathbf{Q}}}{\partial E_{\mathbf{Q}}} \frac{\partial\mu_\text{t}}{\partial n_\text{t}},
    \label{eq:D}
\end{equation}
with the crystal area $A$, the relaxation time $\tau_{\mathbf{Q}}$ (see Appendix\,\ref{SI:diff} for details), the trion group velocity $\mathbf{v}_{\mathbf{Q}}=\frac{\hbar\mathbf{Q}}{M_\text{t}}$, the trion occupation probability in thermal equilibrium $\rho_{\mathbf{Q}}$, the trion chemical potential $\mu_\text{t}$, and the trion density $n_\text{t}=A^{-1}\sum_\mathbf{Q}\rho_\mathbf{Q}$.
In the low density limit where the trion occupation is small, $\rho_\mathbf{Q}$ follows a Boltzmann distribution, the diffusion coefficient then reads\,\cite{rosati2020strain} $D=(2n_\text{t}A)^{-1}\sum_\mathbf{Q}\tau_\mathbf{Q}\mathbf{v}^2_\mathbf{Q}\rho_\mathbf{Q}$, and the relaxation time is given by the out-scattering rate, $\tau_\mathbf{Q}=\Gamma^{-1}_\mathbf{Q}$. We evaluate the trion diffusion coefficient in this regime and obtain approximately 1 cm\textsuperscript{2}/s at temperatures around 5-10 K (cf. Fig.\,\ref{fig4}a).
Compared to excitons (approx. 4.5 cm\textsuperscript{2}/s), trions thus exhibit a considerably slower diffusion reflecting their stronger scattering with phonons and their larger mass (i.e. smaller group velocity).
At higher temperatures, $D$ increases slightly because of the weakening of the scattering rate at the thermally occupied higher energies (cf. Fig\,\ref{fig3}b), reaching 1.6 (4.8) cm\textsuperscript{2}/s for trions (excitons) at 60 K.
We note here that this increase might be counteracted by quantum interference effects\,\cite{glazov2020quantum,wagner2021nonclassical} that are beyond the scope of this work.
Moreover, a larger trion binding energy closer to the experimental values would result in an even less pronounced increase of the diffusion coefficient, although the low-temperature value would remain unchanged as it does not depend on the extension of the trion wave function.
We also note that Eq.\,\eqref{eq:D} can be easily extended to account for multiple valleys. This scenario would be relevant for trion diffusion at higher temperatures, especially in tungsten-based TMDs which exhibit a rich multi-valley band structure\,\cite{rosati2020strain}.

While the calculated diffusion coefficient for excitons agrees well with experimental measurements\,\cite{hotta2020exciton}, the reported value for trions in literature is 4-5 cm\textsuperscript{2}/s at 5 K\,\cite{kim2021free}---i.e. 4-5 times larger than our prediction.
While this discrepancy might be partially explained by the presence of non-equilibrium distributions at such low temperatures, the influence of unpaired electrons or excitons, or perhaps even by the inclusion of electron-hole exchange in the model\,\cite{thompson2022anisotropic}, we show here that the impact of the fermionic nature of trions via Pauli blocking can lead to a drastic enhancement of the diffusion coefficient at low temperatures.
We now consider large trion occupations where $\rho_\mathbf{Q}$ follows a Fermi-Dirac distribution. In this scenario, the relaxation time is not given by the out-scattering rate anymore, but by the sum of in- and out-scattering rates including Pauli blocking\,\cite{hess1996maxwell} (see Appendix\,\ref{SI:diff}). More importantly, the filling of states with low kinetic energy will force the occupation of higher states at elevated densities, thus resulting in an enhanced propagation as the higher states have a larger group velocity. Indeed,  we find a drastic increase from 1 cm\textsuperscript{2}/s in the low-density limit to 4 cm\textsuperscript{2}/s for $n_\text{t}=4 \times 10^{11}$ cm\textsuperscript{-2} at $T = 5$ K. This effect is more prominent at low temperatures, where the trion occupation is higher.
At even larger densities, we expect trion-trion interactions to play an important role via repulsive drift\,\cite{cheng2021observation} and by creating additional scattering channels.

The enhancement of diffusion due to Pauli blocking is a result of the elevated pressure of the degenerate Fermi gas formed by trions. The current density at $T=0$ K  reads (assuming a constant $\tau_\mathbf{Q}=\tau$) $\mathbf{j}=-\frac{\tau}{M_\text{t}}\mathbf{\nabla}P$, where $P=\frac{1}{2}n_\text{t} E_\text{F}$ is the pressure of a 2D degenerate Fermi gas with Fermi energy $E_\text{F}$ (see Appendix\,\ref{SI:diff}). Thus, trion diffusion is driven by a pressure gradient. This is true also for small occupations, where the pressure instead follows the ideal gas law, $P = n_\text{t} k_B T$. Note that we can recover the first Fick's law, $\mathbf{j}=-D\mathbf{\nabla}n_\text{t}$, where $D=\frac{\tau}{M_\text{t}}E_\text{F}$ for the degenerate trion gas at 0 K and $D=\frac{\tau}{M_\text{t}}k_BT$ for the non-degenerate (i.e. low density) gas at finite $T$. This expression clearly shows that the diffusion coefficient should increase linearly with the Fermi level.

Contrary to excitons, trions have an electric charge and can therefore generate a current in the presence of an electric field. This has important technological implications, e.g. the conductivity $\sigma$ in photoexcited doped materials can have a significant contribution from trions\,\cite{lui2014trion}.
The capability of these three-particle charge carriers to be accelerated by the electric field is given by the trion mobility $\tilde{\mu}_\text{t}=\sigma_\text{t}/n_\text{t}$, which is related to the diffusion coefficient via\,\cite{hess1996maxwell} $\tilde{\mu}_\text{t}=D(n_\text{t}\frac{\partial\mu_\text{t}}{\partial n_\text{t}})^{-1}$.
The mobility is a measure for the resistance of a particle to be accelerated by an external potential gradient, and therefore is also meaningful for excitons in, e.g., inhomogeneous strain\,\cite{rosati2021dark} or spatially-varying transverse electric fields\,\cite{unuchek2018room}.
In Fig.\,\ref{fig4}b, we plot the trion mobility for different trion densities and compare it to the exciton and electron mobilities. First of all, we note that the mobility depends very weakly on the density. This is in agreement with the expression $\tilde{\mu}_\text{t}=\tau/M_\text{t}$, which is expected both at small and large occupations assuming a constant $\tau$ (see Appendix\,\ref{SI:diff}). Therefore, the density only influences the mobility via the relaxation time, resulting in a slight increase of $\tilde{\mu}$ due to the Pauli blocking of scattering channels.
Second, the mobility decreases with temperature, following the decrease in the relaxation time, $\tau \propto 1/T$. Finally, we find that the trion mobility is five times (two orders of magnitude) lower than the exciton (electron) mobility.
Concretely, around 10 K we obtain mobilities of $1 \cdot 10^3$, $5 \cdot 10^3$, and $1 \cdot 10^5$ cm\textsuperscript{2}/Vs for trions, excitons, and electrons, respectively.
The low mobility of trions compared to single electrons manifests as a negative differential conductivity in optical-pump THz-probe experiments\,\cite{lui2014trion}.
While the trion mobility has not been directly determined in experiments, the phonon-limited mobility has been estimated to be about $2 \cdot 10^3$ cm\textsuperscript{2}/Vs at low temperatures from linewidth measurements in MoTe\textsubscript{2}\,\cite{helmrich2021high}---in good agreement with our predictions.
Note that defects and dielectric inhomogeneities might play an important role and considerably lower the mobility of TMD samples. Therefore, our  results should be considered as an upper limit for the trion mobility which could be reached in clean samples with a homogeneous dielectric background\cite{raja2019dielectric,zipfel2020exciton}.

\section{Conclusions}
We have developed a microscopic approach to describe trion dynamics in atomically thin semiconductors. The energies and wavefunctions of trions are obtained by solving the trion Schrödinger equation. We resolve the internal quantum structure of bound and continuum trion states and investigate their interaction with phonons for the exemplary case of a MoSe\textsubscript{2} monolayer.
We find that trions exhibit a stronger coupling to phonons compared to excitons and as a result are characterized by a slower spatial propagation. Interestingly, we predict a drastic enhancement of the trion diffusion at low temperatures for increasing densities. This effect is a direct consequence of the fermionic character of trions and can be understood in terms of the large pressure of the degenerate Fermi gas of trions.
Our work provides microscopic insights on trion-phonon interaction and trion propagation, which are interesting from a fundamental perspective but also highly relevant for technological applications.

While we have focused on MoSe\textsubscript{2}, we expect our findings to be qualitatively similar in other atomically-thin semiconductors at low temperatures, although the exact numbers will depend on material-specific parameters such as effective masses, dielectric screening, phonon energies and deformation potentials.
Furthermore, our microscopic approach can be extended to account for electron-hole exchange and to model different aspects of trion dynamics, including valley depolarization, trion-trion interactions, and non-equilibrium trion propagation phenomena analogous to the phonon wind\,\cite{glazov2019phonon} and the Seebeck effect\,\cite{park2021imaging,perea2019exciton} that have been predicted and observed for excitons.
\\[3pt]

\noindent \textbf{Acknowledgements.}
We acknowledge funding from the Chalmers' Excellence Initiative Nano under its Excellence PhD program, the Deutsche Forschungsgemeinschaft
(DFG) via SFB 1083, and the European Union’s Horizon 2020 research and innovation program under grant agreement no. 881603 (Graphene Flagship).
The computations were enabled by resources provided by the Swedish National Infrastructure for Computing (SNIC) at C3SE.

\appendix
\renewcommand{\thesubsection}{\Alph{section}.\arabic{subsection}}
\section{Trion Hamiltonian}
\label{SI:Ht}
In this section, we derive the trion Hamiltonian starting from the electron-hole picture and we introduce the trion Schrödinger equation. The derivation is divided into four parts.
First, we introduce the Hamiltonian in electron-hole basis. Then, we expand the electron and hole creation and annihilation operators in the Fock subspace that contains exactly two electrons and one hole, i.e. a single trion. Then, we introduce the exciton-electron basis and, finally, we transform into a trion basis.

\subsection{Hamiltonian in electron-hole basis}
We consider a system of interacting electrons and holes where electrons can occupy states in the conduction band at the K or K' symmetry points and holes can occupy states in the valence band at the K point (see Fig.\,\ref{fig1}b in the main text). We thus introduce hole and electron annihilation (creation) operators $h^{(\dagger)}_{\mathbf{k}}$, $e^{(\dagger)}_{\mathbf{k}}$, $e'^{(\dagger)}_{\mathbf{k}}$, where $e$ and $e'$ operate on conduction-band states at the K and K' valleys, respectively. We note here that K and K' electrons are distinguishable since they have opposite spin. The Hamilton operator of this system reads
\begin{align}
    H =& \sum_{\mathbf{k}} \left( E^{\text{e}}_{\mathbf{k}} e^{\dagger}_{\mathbf{k}} e^{\phantom{\dagger}}_{\mathbf{k}} + E^{\text{e'}}_{\mathbf{k}} e'^{\dagger}_{\mathbf{k}} e'^{\phantom{\dagger}}_{\mathbf{k}} +  E^{\text{h}}_{\mathbf{k}} h^{\dagger}_{\mathbf{k}} h^{\phantom{\dagger}}_{\mathbf{k}}\right) \nonumber \\
    & + \sum_{\mathbf{kk}\mathbf{'q}} V_{\mathbf{q}} \left( e^{\dagger}_{\mathbf{k}+\mathbf{q}} e'^{\dagger}_{\mathbf{k}'-\mathbf{q}} e'^{\phantom{\dagger}}_{\mathbf{k}'} e^{\phantom{\dagger}}_{\mathbf{k}} - e^{\dagger}_{\mathbf{k}+\mathbf{q}} h^{\dagger}_{\mathbf{k}'-\mathbf{q}} h^{\phantom{\dagger}}_{\mathbf{k}'} e^{\phantom{\dagger}}_{\mathbf{k}} \right. \nonumber \\
    & \left. - e'^{\dagger}_{\mathbf{k}+\mathbf{q}} h^{\dagger}_{\mathbf{k}'-\mathbf{q}} h^{\phantom{\dagger}}_{\mathbf{k}'} e'^{\phantom{\dagger}}_{\mathbf{k}} \right).
    \label{Seq:Hel}
\end{align}
The first line corresponds to quasi-free electrons and holes in the lattice, with $E^{\xi}_{\mathbf{k}}$ being the band energy of the particle species $\xi=\text{e}, \text{e'}, \text{h}$ described in an effective mass approximation with parameters from Ref.\,\cite{kormanyos2015k}. The second and third lines describe the Coulomb interaction between distinguishable electrons and holes with the monolayer potential $V_{\mathbf{q}}$ taken from Ref.\,\cite{van2018coulomb}.
Interactions between indistinguishable particles (e.g. terms like $e^{\dagger}e^{\dagger}e^{\phantom{\dagger}}e^{\phantom{\dagger}}$) have been neglected, since they give rise to scattering and energy renormalization effects that only become relevant at increased trion densities. Moreover, electron-hole exchange, which is known to slightly alter the exciton dispersion and lift the degeneracy between singlet and triplet trions\,\cite{yu2014dirac}, has been disregarded.

\subsection{Expansion of operators in Fock space}
We will now expand our electron and hole creation and annihilation operators in terms of a three-particle (K electron, K' electron, K hole) operator. For this purpose, we have taken the approach used for excitons in Refs.\,\cite{ivanov1993self,katsch2018theory} and extended it to describe trions. The unit operator in Fock space reads,
\begin{align}
    \mathds{1} =& \ket{0}\bra{0} \nonumber \\
    & + \sum_{\mathbf{k}} \left( e^{\dagger}_{\mathbf{k}} \ket{0}\bra{0} e^{\phantom{\dagger}}_{\mathbf{k}} + e'^{\dagger}_{\mathbf{k}} \ket{0}\bra{0} e'^{\phantom{\dagger}}_{\mathbf{k}} + h^{\dagger}_{\mathbf{k}} \ket{0}\bra{0} h^{\phantom{\dagger}}_{\mathbf{k}} \right) \nonumber \\
    & + \sum_{\mathbf{k}_1\mathbf{k}_2} \left( e^{\dagger}_{\mathbf{k}_1} e'^{\dagger}_{\mathbf{k}_2} \ket{0}\bra{0} e'^{\phantom{\dagger}}_{\mathbf{k}_2} e^{\phantom{\dagger}}_{\mathbf{k}_1}   +   e^{\dagger}_{\mathbf{k}_1} h^{\dagger}_{\mathbf{k}_2} \ket{0}\bra{0} h^{\phantom{\dagger}}_{\mathbf{k}_2} e^{\phantom{\dagger}}_{\mathbf{k}_1} \right. \nonumber \\
    & \left. + e'^{\dagger}_{\mathbf{k}_1} h^{\dagger}_{\mathbf{k}_2} \ket{0}\bra{0} h^{\phantom{\dagger}}_{\mathbf{k}_2} e'^{\phantom{\dagger}}_{\mathbf{k}_1}\right) + ...
\end{align}
Here $\ket{0}$ denotes the ground state where the valence band is fully occupied and the conduction band is empty. We have written in the expression above only the terms that we will need later. The remaining terms are irrelevant in our case, as they are only important at high densities or in the presence of unpaired particles. We will later demand that the number of K electrons, K' electrons, and K holes in the system is equal. This assumption allows us to formulate a trion Hamiltonian and provide insights into the physics of trions.

We will introduce the unit operator between creation and annihilation operators in order to write our Hamiltonian in terms of three-particle electron-electron-hole (e-e'-h) operators, $\hat{T}^{\dagger}_{\mathbf{k}_1\mathbf{k}_2\mathbf{k}_3} = e^{\dagger}_{\mathbf{k}_1} e'^{\dagger}_{\mathbf{k}_2} h^{\dagger}_{\mathbf{k}_3}$. We provide here two examples of this expansion:
\begin{align}
    e^{\dagger}_{\mathbf{k}_1} \mathds{1} e^{\phantom{\dagger}}_{\mathbf{k}_2} &= \sum_{\mathbf{k}_3\mathbf{k}_4} e^{\dagger}_{\mathbf{k}_1} e'^{\dagger}_{\mathbf{k}_3} h^{\dagger}_{\mathbf{k}_4} \ket{0}\bra{0} h^{\phantom{\dagger}}_{\mathbf{k}_4} e'^{\phantom{\dagger}}_{\mathbf{k}_3} e^{\phantom{\dagger}}_{\mathbf{k}_2} \nonumber \\
    & = \sum_{\mathbf{k}_3\mathbf{k}_4} \hat{T}^{\dagger}_{\mathbf{k}_1\mathbf{k}_3\mathbf{k}_4} \hat{T}^{\phantom{\dagger}}_{\mathbf{k}_2\mathbf{k}_3\mathbf{k}_4}, \\
    e^{\dagger}_{\mathbf{k}_1} e'^{\dagger}_{\mathbf{k}_2} \mathds{1} e'^{\phantom{\dagger}}_{\mathbf{k}_3} e^{\phantom{\dagger}}_{\mathbf{k}_4} &= \sum_{\mathbf{k}_5} e^{\dagger}_{\mathbf{k}_1} e'^{\dagger}_{\mathbf{k}_2} h^{\dagger}_{\mathbf{k}_5} \ket{0}\bra{0} h^{\phantom{\dagger}}_{\mathbf{k}_5} e'^{\phantom{\dagger}}_{\mathbf{k}_3} e^{\phantom{\dagger}}_{\mathbf{k}_4} \nonumber \\
    & = \sum_{\mathbf{k}_5} \hat{T}^{\dagger}_{\mathbf{k}_1\mathbf{k}_2\mathbf{k}_5} \hat{T}^{\phantom{\dagger}}_{\mathbf{k}_4\mathbf{k}_3\mathbf{k}_5}.
\end{align}
Note that, first, we have only considered terms of the form $e^{\dagger}e'^{\dagger}h^{\dagger}\ket{0}\bra{0}he'e$ which fulfill our restriction that there are exactly three different particles, and second, we have used $\ket{0}\bra{0} = \mathds{1} - ...$\,\cite{katsch2018theory} and disregarded the terms beyond $\mathds{1}$ that would describe trion-trion interactions and are only relevant at high densities.

\subsection{Hamiltonian in exciton-electron basis}
Now, we consider the Hamiltonian in the electron-hole picture, Eq.\,\eqref{Seq:Hel}, and apply the expansion described in the previous section. We obtain
\begin{align}
    H =& \sum_{\mathbf{k}_1\mathbf{k}_2\mathbf{k}_3} \left( E^{\text{e}}_{\mathbf{k}_1} + E^{\text{e'}}_{\mathbf{k}_2} +  E^{\text{h}}_{\mathbf{k}_3} \right) \hat{T}^{\dagger}_{\mathbf{k}_1\mathbf{k}_2\mathbf{k}_3} \hat{T}^{\phantom{\dagger}}_{\mathbf{k}_1\mathbf{k}_2\mathbf{k}_3} \nonumber \\
    &+ \sum_{\mathbf{k}_1\mathbf{k}_2\mathbf{k}_3\mathbf{q}} V_{\mathbf{q}} \left( \hat{T}^{\dagger}_{\mathbf{k}_1+\mathbf{q},\mathbf{k}_2-\mathbf{q},\mathbf{k}_3} - \hat{T}^{\dagger}_{\mathbf{k}_1+\mathbf{q},\mathbf{k}_2,\mathbf{k}_3-\mathbf{q}} \right. \nonumber \\
    & \left. - \hat{T}^{\dagger}_{\mathbf{k}_1,\mathbf{k}_2+\mathbf{q},\mathbf{k}_3-\mathbf{q}} \right) \hat{T}^{\phantom{\dagger}}_{\mathbf{k}_1\mathbf{k}_2\mathbf{k}_3}.
    \label{Seq:Heeh}
\end{align}
In principle, we could now diagonalize the Hamiltonian by expanding $\hat{T}^{\dagger}_{\mathbf{k}_1\mathbf{k}_2\mathbf{k}_3}$ in the eigenbasis of the three-body Schrödinger equation associated to the considered trion. However, it is advantageous to first expand it in the exciton eigenbasis\,\cite{kira2006many,berghauser2018mapping}, in which a part of the Hamiltonian containing the energies of the K electron (e) and the K hole (h) as well as their Coulomb interaction becomes diagonal. Thus, we write
\begin{equation}
    \hat{T}^{\dagger}_{\mathbf{k}_1\mathbf{k}_2\mathbf{k}_3} = \sum_{\nu} \phi^{\nu}_{\alpha_\text{h} \mathbf{k}_1 - \alpha_\text{e} \mathbf{k}_3} \tilde{T}^{\dagger}_{\nu,\mathbf{k}_1+\mathbf{k}_3,\mathbf{k}_2},
\end{equation}
where $\alpha_{\xi}=m_\xi/M_\text{x}$ and $M_\text{x}=m_\text{e}+m_\text{h}$. $\phi^{\nu}_{\mathbf{k}_\text{x}}$ is the wave function of the exciton state $\nu$ with electron-hole relative momentum $\mathbf{k}_\text{x}$ and fulfills the Wannier equation,
\begin{equation}
    \frac{\hbar^2\mathbf{k}_\text{x}^2}{2m_\text{x}}\phi^{\nu}_{\mathbf{k}_\text{x}} - \sum_{\mathbf{q}}V_{\mathbf{q}}\phi^{\nu}_{\mathbf{k}_\text{x}+\mathbf{q}}=\varepsilon_\text{x}^{\nu} \phi^{\nu}_{\mathbf{k}_\text{x}},
\end{equation}
with $m_\text{x}=m_\text{e}m_\text{h}/M_\text{x}$ being the exciton reduced mass and $\varepsilon_\text{x}^{\nu}$ the exciton binding energy.
We have also introduced the exciton-electron creation operator $\tilde{T}^{\dagger}_{\nu,\mathbf{Q}_\text{x},\mathbf{k}_\text{e}}$, which creates an exciton at state $\nu$ with electron-hole center-of-mass momentum $\mathbf{Q}_\text{x}$ and an electron with momentum $\mathbf{k}_\text{e}$.
Note that one can also express the exciton-electron operator as $\tilde{T}^{\dagger}_{\nu,\mathbf{Q}_\text{x},\mathbf{k}_\text{e}} = X^{\dagger}_{\nu,\mathbf{Q}_\text{x}}e'^{\dagger}_{\mathbf{k}_\text{e}}$, where $X^{\dagger}_{\nu,\mathbf{Q}_\text{x}}$ is the exciton creation operator.

Applying the expansion into exciton basis to Eq.\,\eqref{Seq:Heeh} leads to the exciton-electron Hamiltonian,
\begin{align}
    H =& \sum_{\mathbf{\mathbf{Q}_\text{x}\mathbf{k}_\text{e}}\nu} \left[ \left( \varepsilon_\text{x}^{\nu} + \frac{\hbar^2\mathbf{Q}_\text{x}^2}{2M_\text{x}} + \frac{\hbar^2\mathbf{k}_\text{e}^2}{2m_\text{e}} \right) \tilde{T}^{\dagger}_{\nu,\mathbf{Q}_\text{x},\mathbf{k}_\text{e}} \right. \nonumber \\
    & \left. + \sum_{\mu\mathbf{q}} \tilde{V}^{\nu\mu}_{\mathbf{q}} \tilde{T}^{\dagger}_{\mu,\mathbf{Q}_\text{x}+\mathbf{q},\mathbf{k}_\text{e}-\mathbf{q}} \right] \tilde{T}^{\phantom{\dagger}}_{\nu,\mathbf{Q}_\text{x},\mathbf{k}_\text{e}}.
    \label{Seq:Hxe}
\end{align}
The interaction between electrons and holes appears now as an exciton-electron interaction with the matrix element
\begin{align}
    \tilde{V}^{\nu\mu}_{\mathbf{q}} &= V_{\mathbf{q}}\braket{\nu|\left(\mathrm{e}^{i\alpha_\text{h}\mathbf{q \cdot r}}-\mathrm{e}^{-i\alpha_\text{e}\mathbf{q \cdot r}}\right)|\mu} \nonumber \\
    & = V_{\mathbf{q}}\sum_{\mathbf{k}_\text{x}} \phi^{\nu *}_{\mathbf{k}_\text{x}} \left( \phi^{\mu}_{\mathbf{k}_\text{x}+\alpha_\text{h}\mathbf{q}} - \phi^{\mu}_{\mathbf{k}_\text{x}-\alpha_\text{e}\mathbf{q}} \right).
\end{align}

\subsection{Hamiltonian in trion basis}
We now expand the exciton-electron operators in trion basis, i.e.
\begin{equation}
    \tilde{T}^{\dagger}_{\nu,\mathbf{Q}_\text{x},\mathbf{k}_\text{e}} = \sum_{\lambda} \psi^{\lambda}_{\nu,\beta_\text{e}\mathbf{Q}_\text{x}-\beta_\text{x}\mathbf{k}_\text{e}} T^{\dagger}_{\lambda,\mathbf{Q}_\text{x}+\mathbf{k}_\text{e}},
\end{equation}
with $\beta_\text{e}=m_\text{e}/M_\text{t}, \beta_\text{x}=M_\text{x}/M_\text{t}$, where $M_\text{t}=M_\text{x}+m_\text{e}$ is the trion total mass. Here $T^{\dagger}_{\lambda,\mathbf{Q}_\text{t}}$ creates a trion at state $\lambda$ with center-of-mass momentum $\mathbf{Q}_\text{t}$, and $\psi^{\lambda}_{\nu,\mathbf{k}_\text{t}}$ is the wave function of the trion state $\lambda$ with exciton-electron relative momentum $\mathbf{k}_\text{t}$ and exciton state $\nu$. This wave function fulfills the following trion Schrödinger equation,
\begin{equation}
    \left( \varepsilon_\text{x}^{\nu}+\frac{\hbar^2\mathbf{k}_\text{t}^2}{2m_\text{x-e}} \right) \psi^{\lambda}_{\nu,\mathbf{k}_\text{t}} + \sum_{\mu\mathbf{q}} \tilde{V}^{\nu\mu}_{\mathbf{q}} \psi^{\lambda}_{\mu,\mathbf{k}_\text{t}+\mathbf{q}} = \varepsilon_\text{t}^{\lambda} \psi^{\lambda}_{\nu,\mathbf{k}_\text{t}},
    \label{Seq:trionSchrodinger}
\end{equation}
where $m_\text{x-e}=M_\text{x}m_\text{e}/M_\text{t}$ is the exciton-electron reduced mass and $\varepsilon_\text{t}^{\lambda}$ is the trion eigenenergy.
Writing Eq.\,\eqref{Seq:Hxe} in trion basis finally leads to the trion Hamiltonian,
\begin{equation}
    H = \sum_{\lambda\mathbf{Q}_\text{t}} \left( \varepsilon_\text{t}^{\lambda} + \frac{\hbar^2\mathbf{Q}_\text{t}^2}{2M_\text{t}} \right) T^{\dagger}_{\lambda,\mathbf{Q}_\text{t}} T^{\phantom{\dagger}}_{\lambda,\mathbf{Q}_\text{t}}.
\end{equation}
We note that the trion operators fulfill fermionic anti-commutation relations in the low-density limit,
\begin{align}
    \left\{ T^{\phantom{\dagger}}_{\lambda,\mathbf{Q}_\text{t}}, T^{\dagger}_{\lambda',\mathbf{Q}_\text{t}'} \right\} &= \delta_{\lambda,\lambda'}\delta_{\mathbf{Q}_\text{t},\mathbf{Q}_\text{t}'} \\  \left\{ T^{(\dagger)}_{\lambda,\mathbf{Q}_\text{t}}, T^{(\dagger)}_{\lambda',\mathbf{Q}_\text{t}'} \right\} &= 0.
\end{align}
In principle, the anti-commutator $\{ T^{\phantom{\dagger}}_{\lambda,\mathbf{Q}_\text{t}}, T^{\dagger}_{\lambda',\mathbf{Q}_\text{t}'} \}$ should also contain corrections of the form $T^{\dagger}T$ which would be relevant at high densities and would influence trion-trion scattering and energy renormalization. Since we consider sufficiently low densities throughout this work, we can disregard these corrections and treat trions as purely fermionic quasi-particles.

\section{Trion-phonon Hamiltonian}
\label{SI:Htp}
The Hamilton operator describing the interaction between acoustic phonons and the electron and hole species considered here reads
\begin{align}
    H_\text{t-p} =& \sum_{\mathbf{kq}} \left( g^\text{e}_{\mathbf{q}}e^{\dagger}_{\mathbf{k}+\mathbf{q}}e^{\phantom{\dagger}}_{\mathbf{k}} + g^\text{e'}_{\mathbf{q}}e'^{\dagger}_{\mathbf{k}+\mathbf{q}}e'^{\phantom{\dagger}}_{\mathbf{k}} + g^\text{h}_{\mathbf{q}}h^{\dagger}_{\mathbf{k}+\mathbf{q}}h^{\phantom{\dagger}}_{\mathbf{k}} \right) \nonumber \\
    & \times \left( b^{\phantom{\dagger}}_{\mathbf{q}} + b^{\dagger}_{-\mathbf{q}} \right),
\end{align}
where $g^\xi_\mathbf{q} = \sqrt{\frac{\hbar}{2\rho\Omega_{\mathbf{q}}A}}D^{\xi}_{\mathbf{q}}$ is the electron-phonon matrix element, $\rho$ is the material's mass density, $\Omega_{\mathbf{q}}$ is the phonon frequency, $A$ is the crystal area, and $D^\xi_\mathbf{q}$ is the deformation potential.
Note that the phonon coupling for holes is related to that of valence band electrons via $g^\text{h}_\textbf{q}=-g^\text{v}_\textbf{q}$.
In principle, a phonon mode index should appear in the Hamiltonian above. However, we consider here only one effective acoustic phonon mode.
For long-wavelength acoustic phonons, $D^\xi_\mathbf{q}=\tilde{D}^\xi|\mathbf{q}|$, with $\tilde{D}^{\xi}$ exhibiting the same sign for electrons and holes\,\cite{peelaers2012effects} and thus reflecting the non-polar nature of the interaction. In particular, we consider $\tilde{D}^\text{e}\approx\tilde{D}^\text{h}$\,\cite{jin2014intrinsic} and obtain $|\tilde{D}^\text{e}|=1.4$ eV by fitting\,\cite{shree2018observation} the acoustic phonon contribution to the experimentally measured exciton linewidth\,\cite{hotta2020exciton}. The acoustic phonon frequency is given by $\Omega_\mathbf{q}=s|\mathbf{q}|$ where $s=4.1 \times 10^5\ \text{cm}/\text{s}$\,\cite{jin2014intrinsic}.

In order to write this Hamiltonian in trion basis, we proceed as in Section\,\ref{SI:Ht}. First, we transform into exciton basis and obtain the trion-phonon Hamiltonian in the exciton-electron picture,
\begin{align}
    H_\text{t-p} =& \sum_{\mathbf{Q}_\text{x}\mathbf{k}_\text{e}\mathbf{q}\nu\mu} \left( G^{\nu\mu}_{\text{x},\mathbf{q}}\tilde{T}^{\dagger}_{\nu,\mathbf{Q}_\text{x}+\mathbf{q},\mathbf{k}_\text{e}} + \delta_{\nu\mu} g^\text{e}_{\mathbf{q}}\tilde{T}^{\dagger}_{\nu,\mathbf{Q}_\text{x},\mathbf{k}_\text{e}+\mathbf{q}} \right) \nonumber \\
    & \times \tilde{T}^{\phantom{\dagger}}_{\mu,\mathbf{Q}_\text{x},\mathbf{k}_\text{e}} \left( b^{\phantom{\dagger}}_{\mathbf{q}} + b^{\dagger}_{-\mathbf{q}} \right)
\end{align}
Here $G^{\nu\mu}_{\text{x},\mathbf{q}}=\sum_{\mathbf{k}_\text{x}}\phi^{\mu *}_{\mathbf{k}_\text{x}} \left( g^\text{e}_\mathbf{q}\phi^\nu_{\mathbf{k_\text{x}}+\alpha_\text{h}\mathbf{q}} + g^\text{h}_\mathbf{q}\phi^\nu_{\mathbf{k_\text{x}}-\alpha_\text{e}\mathbf{q}} \right)$ is the exciton-phonon matrix element\,\cite{selig16}. Transforming the exciton-electron operators into trion basis leads to the trion-phonon Hamiltonian,
\begin{equation}
    H_\text{t-p} = \sum_{\mathbf{Q}_\text{t}\mathbf{q}\lambda\lambda'} G^{\lambda\lambda'}_{\text{t},\mathbf{q}} T^{\dagger}_{\lambda,\mathbf{Q}_\text{t}+\mathbf{q}} T^{\phantom{\dagger}}_{\lambda',\mathbf{Q}_\text{t}} \left( b^{\phantom{\dagger}}_{\mathbf{q}} + b^{\dagger}_{-\mathbf{q}} \right),
\end{equation}
with the trion-phonon matrix element,
\begin{equation}
    G^{\lambda\lambda'}_{\text{t},\mathbf{q}} = \sum_{\mathbf{k}_\text{t}\nu\mu} \psi^{\lambda'*}_{\mu,\mathbf{k}_\text{t}} \left(  G^{\nu\mu}_{\text{x},\mathbf{q}} \psi^\lambda_{\nu,\mathbf{k}_\text{t}+\beta_\text{e}\mathbf{q}} + \delta_{\nu\mu}g^\text{e}_\mathbf{q} \psi^\lambda_{\nu,\mathbf{k}_\text{t}-\beta_\text{x}\mathbf{q}} \right).
\end{equation}
This matrix element has two terms. The first one corresponds to exciton-phonon interaction and the second one describes electron-phonon interaction. Note that for the electron term, the exciton indices $\nu$ and $\mu$ must be the same, i.e. no excitonic transition occurs.

\section{Quasi-1D trion Schrödinger equation}
\label{SI:1D}
The trion Schrödinger equation\,\eqref{Seq:trionSchrodinger} is numerically too demanding to be solved. Nevertheless, similar to the Wannier equation, its complexity can be significantly reduced by eliminating the angular degree of freedom of the relative momentum. We consider the following ansatz for the trion wave function,
\begin{equation}
    \psi^{\lambda}_{\nu,\mathbf{k}} = \mathrm{e}^{i L_{\lambda\nu} \theta_{\mathbf{k}}} \tilde{\psi}^{\lambda}_{\nu,|\mathbf{k}|},
\end{equation}
allowing us to separate the radial and angular momentum coordinates.
A similar approach has been taken in Ref.\,\cite{katsch2022excitonic}, where the wave function has been expanded in a Fourier series in angular coordinates.
We introduce our ansatz in Eq.\,\eqref{Seq:trionSchrodinger}, multiply by $\mathrm{e}^{-i L_{\lambda\nu} \theta_{\mathbf{k}}}$, integrate over $\theta_{\mathbf{k}}$, and rename $\mathbf{q}=\mathbf{k}'-\mathbf{k}$, obtaining
\begin{align}
    & \left( \varepsilon_\text{x}^{\nu}+\frac{\hbar^2|\mathbf{k}|^2}{2m_\text{x-e}} \right) \tilde{\psi}^{\lambda}_{\nu,|\mathbf{k}|} \nonumber \\
    & + \frac{1}{2\pi} \int_0^{2\pi}d\theta_{\mathbf{k}} \sum_{\mu\mathbf{k}'} \tilde{V}^{\nu\mu}_{\mathbf{k}'-\mathbf{k}} \mathrm{e}^{-iL_{\lambda\nu}\theta_{\mathbf{k}}} \mathrm{e}^{iL_{\lambda\mu}\theta_{\mathbf{k}'}} \tilde{\psi}^{\lambda}_{\mu,|\mathbf{k}'|} \nonumber \\
    & = \varepsilon_\text{t}^{\lambda} \tilde{\psi}^{\lambda}_{\nu,|\mathbf{k}|}.
    \label{Seq:trionSchrodinger_step1}
\end{align}
In order to evaluate the integral over $\theta_{\mathbf{k}}$, we first note that
$
    \tilde{V}^{\nu\mu}_{\mathbf{q}} = \mathrm{e}^{i(m_\mu-m_\nu)\theta_{\mathbf{q}}} \tilde{V}^{\nu\mu}_{|\mathbf{q}|\hat{x}},
$
which can be found by using the separation of variables for the exciton wave function, $\phi^{\nu}_{\mathbf{k}} = \mathrm{e}^{i m_\nu \theta_{\mathbf{k}}} \tilde{\phi}^{\nu}_{|\mathbf{k}|}$, with $m_\nu$ being the exciton angular quantum number. We further note that
$
    |\mathbf{k}'-\mathbf{k}|\mathrm{e}^{i\theta_{\mathbf{k}'-\mathbf{k}}} = |\mathbf{k}'|\mathrm{e}^{i\theta_{\mathbf{k}'}} - |\mathbf{k}|\mathrm{e}^{i\theta_{\mathbf{k}}}.
$
Introducing these in Eq.\,\eqref{Seq:trionSchrodinger_step1}, shifting $\theta_{\mathbf{k}'} \rightarrow \theta_{\mathbf{k}'} + \theta_{\mathbf{k}}$, and evaluating the integral over $\theta_{\mathbf{k}}$ yields a non-zero term only when $L_{\lambda\nu}-L_{\lambda\mu}=m_\mu-m_\nu$. Hence, the separation ansatz $L_{\lambda\nu}=l_\lambda+\tilde{m}_\nu$ yields $\tilde{m}_\nu-\tilde{m}_\mu=m_\mu-m_\nu$, which is fulfilled if $\tilde{m}_\nu=-m_\nu$.
After evaluating the $\theta_{\mathbf{k}}$ integral, the trion Schrödinger equation becomes effectively one-dimensional, in the sense that it depends on the radial component of $\mathbf{k}$ but not on its direction. The trion Schrödinger equation now reads
\begin{align}
    & \left( \varepsilon_\text{x}^{\nu}+\frac{\hbar^2|\mathbf{k}|^2}{2m_\text{x-e}} \right) \tilde{\psi}^{\lambda}_{\nu,|\mathbf{k}|} \nonumber \\
    & + \sum_{\mu\mathbf{k}'} \tilde{V}^{\nu\mu}_{|\mathbf{k}'-\mathbf{k}|\hat{x}} \left[F(|\mathbf{k}|,|\mathbf{k}'|,\theta_{\mathbf{k}'})\right]^{(m_\mu-m_\nu)} \mathrm{e}^{i(l_\lambda-m_\mu)\theta_{\mathbf{k}'}} \tilde{\psi}^{\lambda}_{\mu,|\mathbf{k}'|} \nonumber \\
    & = \varepsilon_\text{t}^{\lambda} \tilde{\psi}^{\lambda}_{\nu,|\mathbf{k}|},
    \label{Seq:trionSchrodinger_effective}
\end{align}
where
$$
F(|\mathbf{k}|,|\mathbf{k}'|,\theta_{\mathbf{k}'}) = \frac{|\mathbf{k}'|\mathrm{e}^{i\theta_{\mathbf{k}'}}-|\mathbf{k}|}{\sqrt{|\mathbf{k}|^2+|\mathbf{k}'|^2-2|\mathbf{k}||\mathbf{k}'|\mathrm{cos}\theta_{\mathbf{k}'}}}.
$$
We numerically solve Eq.\,\eqref{Seq:trionSchrodinger_effective} for $l_\lambda=0$ (s-type trion)  considering 7 bound exciton states and the exciton continuum with $m_\nu=0,\pm 1,\pm 2,\pm 3$. This large amount of states is necessary to obtain a converged trion binding energy.

In order to model a realistic hBN-encapsulated MoSe\textsubscript{2} monolayer, we consider the effective masses\,\cite{kormanyos2015k} $m_\text{e}=0.5m_0, m_\text{h}=0.6m_0$, as well as the material-specific Coulomb potential from Ref.\,\cite{van2018coulomb} with TMD thickness $d=0.6\ \text{nm}$ and polarizability $\chi=7.1d/(2\pi)$, along with the hBN dielectric constant\,\cite{geick1966normal} $\epsilon_\text{hBN}=4.5$.

\section{Phonon-induced trion cooling}
\label{SI:cooling}
In the main text, we evaluate trion cooling times. Here, we present the model that describes the cooling due to trion-phonon scattering. The quantity that describes energy dissipation is the cooling power, $Q=A^{-1}\sum_\mathbf{q}\hbar\Omega_\mathbf{q}\dot{n}_\mathbf{q}$, i.e. the rate at which the trions transfer their excess energy into the thermal bath of phonons. Here, $n_\mathbf{q}=\braket{b^{\dagger}_\mathbf{q}b^{\phantom{\dagger}}_\mathbf{q}}$ is the phonon number. We follow the derivation in Ref.\,\cite{kaasbjerg2014hot} and find
\begin{align}
    Q =& \frac{2\pi}{\hbar A} \sum_{\mathbf{Q}\mathbf{q}}\hbar\Omega_\mathbf{q}|G_{\text{t},\mathbf{q}}|^2 \left[n_\mathbf{q}(T)-n_\mathbf{q}(T_\text{L})\right] \nonumber \\
    & \times \left[\rho_\mathbf{Q}(T)-\rho_{\mathbf{Q}+\mathbf{q}}(T)\right]\delta\!\left(E_{\mathbf{Q}+\mathbf{q}}-E_\mathbf{Q}-\hbar\Omega_{\mathbf{q}}\right),
    \label{eq:cooling}
\end{align}
where $T$ and $T_\text{L}$ are the trion and lattice temperatures, respectively, $\rho_\mathbf{Q}$ is the trion occupation which we assume to be in thermal equilibrium, and $E_\mathbf{Q}=\hbar^2\mathbf{Q}^2/(2M_\text{t})$ is the trion kinetic energy. Here we have dropped the trion state index $\lambda$, since we consider only the trion ground state.
At sufficiently low densities, high temperatures, and disregarding the impact of trion and exciton wave function overlaps, one recovers the expression from Ref.\cite{zipfel2022electron}, i.e. $Q/n_\text{t}=\tau_\text{c}^{-1}k_\text{B}(T-T_\text{L})$, with the cooling time $\tau_\text{c}^{-1}=2m^2|2D^\text{e}+D^\text{h}|^2/(\rho\hbar^3)$ and the trion density $n_\text{t}$. Here, instead, we evaluate Eq.\,\eqref{eq:cooling} without further approximations for temperatures ranging from 5 to 50 K and densities up to $4 \cdot 10^{11}\ \text{cm}^{-2}$ and compute the effective cooling time as $\tau_\text{c}=n_\text{t}k_\text{B}(T-T_\text{L})/Q$. We obtain values in the range of 4-10 ps, in good agreement with experimental measurements\,\cite{zipfel2022electron}.

\section{Diffusion coefficient and mobility}
\label{SI:diff}
We derive equations for the diffusion coefficient $D$ and mobility $\tilde{\mu}$ of trions following Ref.\,\cite{hess1996maxwell} and obtain
\begin{align}
    D &= -\frac{1}{2A}\sum_{\mathbf{Q}} \tau_{\mathbf{Q}} \mathbf{v}_{\mathbf{Q}}^2 \frac{\partial\rho_{\mathbf{Q}}}{\partial E_{\mathbf{Q}}} \frac{\partial\mu}{\partial n_\text{t}}, \\
    \tilde{\mu} &= -\frac{1}{2An_\text{t}}\sum_{\mathbf{Q}} \tau_{\mathbf{Q}} \mathbf{v}_{\mathbf{Q}}^2 \frac{\partial\rho_{\mathbf{Q}}}{\partial E_{\mathbf{Q}}},
\end{align}
with the crystal area $A$, the relaxation time $\tau_{\mathbf{Q}}$, the trion group velocity $\mathbf{v}_{\mathbf{Q}}=\hbar\mathbf{Q}/M_\text{t}$, the trion occupation in thermal equilibrium $\rho_{\mathbf{Q}}$, the trion chemical potential $\mu$, and the trion density $n_\text{t}=A^{-1}\sum_\mathbf{Q}\rho_\mathbf{Q}$.
We introduce here the relaxation time, which does not appear in the main text,
\begin{equation}
    \tau^{-1}_\mathbf{Q} = \sum_{\mathbf{Q}'} \left[ \Gamma_{\mathbf{Q}\mathbf{Q}'}\rho_{\mathbf{Q}'} + \Gamma_{\mathbf{QQ}'} \left(1-\rho_{\mathbf{Q}}\right) \right],
\end{equation}
where
$$
\Gamma_{\mathbf{QQ}'} = \sum_\pm \frac{2\pi}{\hbar}|G_{\text{t},\mathbf{q}}|^2 \eta^\pm_\mathbf{q} \delta\!\left(E_{\mathbf{Q}'}-E_\mathbf{Q}\pm\hbar\Omega_\mathbf{q}\right)
$$
is the scattering matrix with $\mathbf{q}=\mathbf{Q}'-\mathbf{Q}$ and $\eta^\pm_{\mathbf{q}}=n_{\mathbf{q}}+\frac{1}{2}\pm\frac{1}{2}$.
We evaluate the equations for $D$ and $\tilde{\mu}$ without further approximation and plot the results in Fig.\,\ref{fig4} in the main text. In addition, in order to get a better understanding of trion diffusion and mobility, we discuss how $D$ and $\tilde{\mu}$ behave in limiting cases.
At low occupations, the relaxation time reads $\tau^{-1}_\mathbf{Q}=\sum_{\mathbf{Q}'}\Gamma_{\mathbf{QQ}'}$, which is the out-scattering rate that we have introduced in the main text.

We now derive the diffusion coefficient at $T=0\ \text{K}$ to show the effect of the degeneracy pressure. For this purpose, it is convenient to recover the expression for the current and assume $\tau_\mathbf{Q}=\tau$. We thus have $\mathbf{j}(\mathbf{r})=-\tau/(2A)\sum_\mathbf{Q}\mathbf{v}^2_\mathbf{Q}\mathbf{\nabla}\rho_\mathbf{Q}(\mathbf{r})$. At 0 K the trion distribution follows $\rho_\mathbf{Q}=1$ if $|\mathbf{Q}|<Q_\text{F}$ or $0$ otherwise, where $Q_\text{F}=2\sqrt{\pi n_\text{t}}$ is the radius of the 2D Fermi sphere.
We then obtain
\begin{align}
    \mathbf{j}(\mathbf{r}) &= -\frac{\tau\hbar^2}{16\pi M_\text{t}^2}\mathbf{\nabla}\left[\mathbf{Q}^4_\text{F}(\mathbf{r})\right] =-\frac{\tau}{2M_\text{t}}\mathbf{\nabla}\left[n_\text{t}(\mathbf{r})E_\text{F}(\mathbf{r})\right] \nonumber \\
    &= -\frac{\tau}{M_\text{t}}\mathbf{\nabla}P(\mathbf{r}),
\end{align}
where in the last step we have identified the pressure of a 2D degenerate Fermi gas. The pressure is calculated with $P=-\partial U/\partial A$ keeping the total number of particles fixed, where $U=\sum_{|\mathbf{Q}| \leq Q_\text{F}} \hbar^2\mathbf{Q}^2/(2M_\text{t})$ is the total energy of the trion system.

In the following, we argue that we expect the mobility to be $\tilde{\mu} \approx \tau/M_\text{t}$ at low and high densities, as stated in the main text. As before, we assume $\tau_\mathbf{Q}=\tau$. For high densities, we take the limit $T=0\ \text{K}$ to simplify the problem. Taking $D=E_\text{F}\tau/M_\text{t}$ from our result above and using the relation $\tilde{\mu}=D(\partial E_\text{F}/\partial n_\text{t})^{-1}/n_\text{t}$ we find $\tilde{\mu}=\tau/M_\text{t}$. Thus, we do not expect the mobility to depend on density beyond the density-dependence of $\tau$.
Furthermore, at low densities and finite temperatures, where $\rho_\mathbf{Q}$ is well described with a Boltzmann distribution, the mobility can be simplified to $\tilde{\mu}=\tau\braket{\mathbf{v}^2_\mathbf{Q}}/(2k_\text{B}T)$, where $\braket{\mathbf{v}^2_\mathbf{Q}}=\sum_\mathbf{Q}\mathbf{v}^2_\mathbf{Q}\rho_\mathbf{Q}/(n_\text{t}A)=2k_\text{B}T/M_\text{t}$ following the equipartition principle. Thus we again obtain $\tilde{\mu}=\tau/M_\text{t}$.


\begin{thebibliography}{70}%
\makeatletter
\providecommand \@ifxundefined [1]{%
 \@ifx{#1\undefined}
}%
\providecommand \@ifnum [1]{%
 \ifnum #1\expandafter \@firstoftwo
 \else \expandafter \@secondoftwo
 \fi
}%
\providecommand \@ifx [1]{%
 \ifx #1\expandafter \@firstoftwo
 \else \expandafter \@secondoftwo
 \fi
}%
\providecommand \natexlab [1]{#1}%
\providecommand \enquote  [1]{``#1''}%
\providecommand \bibnamefont  [1]{#1}%
\providecommand \bibfnamefont [1]{#1}%
\providecommand \citenamefont [1]{#1}%
\providecommand \href@noop [0]{\@secondoftwo}%
\providecommand \href [0]{\begingroup \@sanitize@url \@href}%
\providecommand \@href[1]{\@@startlink{#1}\@@href}%
\providecommand \@@href[1]{\endgroup#1\@@endlink}%
\providecommand \@sanitize@url [0]{\catcode `\\12\catcode `\$12\catcode
  `\&12\catcode `\#12\catcode `\^12\catcode `\_12\catcode `\%12\relax}%
\providecommand \@@startlink[1]{}%
\providecommand \@@endlink[0]{}%
\providecommand \url  [0]{\begingroup\@sanitize@url \@url }%
\providecommand \@url [1]{\endgroup\@href {#1}{\urlprefix }}%
\providecommand \urlprefix  [0]{URL }%
\providecommand \Eprint [0]{\href }%
\providecommand \doibase [0]{https://doi.org/}%
\providecommand \selectlanguage [0]{\@gobble}%
\providecommand \bibinfo  [0]{\@secondoftwo}%
\providecommand \bibfield  [0]{\@secondoftwo}%
\providecommand \translation [1]{[#1]}%
\providecommand \BibitemOpen [0]{}%
\providecommand \bibitemStop [0]{}%
\providecommand \bibitemNoStop [0]{.\EOS\space}%
\providecommand \EOS [0]{\spacefactor3000\relax}%
\providecommand \BibitemShut  [1]{\csname bibitem#1\endcsname}%
\let\auto@bib@innerbib\@empty
\bibitem [{\citenamefont {Yu}\ \emph {et~al.}(2015)\citenamefont {Yu},
  \citenamefont {Cui}, \citenamefont {Xu},\ and\ \citenamefont
  {Yao}}]{yu2015valley}%
  \BibitemOpen
  \bibfield  {author} {\bibinfo {author} {\bibfnamefont {H.}~\bibnamefont
  {Yu}}, \bibinfo {author} {\bibfnamefont {X.}~\bibnamefont {Cui}}, \bibinfo
  {author} {\bibfnamefont {X.}~\bibnamefont {Xu}},\ and\ \bibinfo {author}
  {\bibfnamefont {W.}~\bibnamefont {Yao}},\ }\bibfield  {title} {\bibinfo
  {title} {Valley excitons in two-dimensional semiconductors},\ }\href
  {https://doi.org/10.1093/nsr/nwu078} {\bibfield  {journal} {\bibinfo
  {journal} {National Science Review}\ }\textbf {\bibinfo {volume} {2}},\
  \bibinfo {pages} {57} (\bibinfo {year} {2015})}\BibitemShut {NoStop}%
\bibitem [{\citenamefont {Wang}\ \emph {et~al.}(2018)\citenamefont {Wang},
  \citenamefont {Chernikov}, \citenamefont {Glazov}, \citenamefont {Heinz},
  \citenamefont {Marie}, \citenamefont {Amand},\ and\ \citenamefont
  {Urbaszek}}]{wang2018colloquium}%
  \BibitemOpen
  \bibfield  {author} {\bibinfo {author} {\bibfnamefont {G.}~\bibnamefont
  {Wang}}, \bibinfo {author} {\bibfnamefont {A.}~\bibnamefont {Chernikov}},
  \bibinfo {author} {\bibfnamefont {M.~M.}\ \bibnamefont {Glazov}}, \bibinfo
  {author} {\bibfnamefont {T.~F.}\ \bibnamefont {Heinz}}, \bibinfo {author}
  {\bibfnamefont {X.}~\bibnamefont {Marie}}, \bibinfo {author} {\bibfnamefont
  {T.}~\bibnamefont {Amand}},\ and\ \bibinfo {author} {\bibfnamefont
  {B.}~\bibnamefont {Urbaszek}},\ }\bibfield  {title} {\bibinfo {title}
  {Colloquium: Excitons in atomically thin transition metal dichalcogenides},\
  }\href {https://doi.org/10.1103/RevModPhys.90.021001} {\bibfield  {journal}
  {\bibinfo  {journal} {Reviews of Modern Physics}\ }\textbf {\bibinfo {volume}
  {90}},\ \bibinfo {pages} {021001} (\bibinfo {year} {2018})}\BibitemShut
  {NoStop}%
\bibitem [{\citenamefont {Mueller}\ and\ \citenamefont
  {Malic}(2018)}]{mueller2018exciton}%
  \BibitemOpen
  \bibfield  {author} {\bibinfo {author} {\bibfnamefont {T.}~\bibnamefont
  {Mueller}}\ and\ \bibinfo {author} {\bibfnamefont {E.}~\bibnamefont
  {Malic}},\ }\bibfield  {title} {\bibinfo {title} {Exciton physics and device
  application of two-dimensional transition metal dichalcogenide
  semiconductors},\ }\href {https://doi.org/10.1038/s41699-018-0074-2}
  {\bibfield  {journal} {\bibinfo  {journal} {npj 2D Materials and
  Applications}\ }\textbf {\bibinfo {volume} {2}},\ \bibinfo {pages} {1}
  (\bibinfo {year} {2018})}\BibitemShut {NoStop}%
\bibitem [{\citenamefont {Jin}\ \emph {et~al.}(2018)\citenamefont {Jin},
  \citenamefont {Ma}, \citenamefont {Karni}, \citenamefont {Regan},
  \citenamefont {Wang},\ and\ \citenamefont {Heinz}}]{jin2018ultrafast}%
  \BibitemOpen
  \bibfield  {author} {\bibinfo {author} {\bibfnamefont {C.}~\bibnamefont
  {Jin}}, \bibinfo {author} {\bibfnamefont {E.~Y.}\ \bibnamefont {Ma}},
  \bibinfo {author} {\bibfnamefont {O.}~\bibnamefont {Karni}}, \bibinfo
  {author} {\bibfnamefont {E.~C.}\ \bibnamefont {Regan}}, \bibinfo {author}
  {\bibfnamefont {F.}~\bibnamefont {Wang}},\ and\ \bibinfo {author}
  {\bibfnamefont {T.~F.}\ \bibnamefont {Heinz}},\ }\bibfield  {title} {\bibinfo
  {title} {Ultrafast dynamics in van der {W}aals heterostructures},\ }\href
  {https://doi.org/10.1038/s41565-018-0298-5} {\bibfield  {journal} {\bibinfo
  {journal} {Nature Nanotechnology}\ }\textbf {\bibinfo {volume} {13}},\
  \bibinfo {pages} {994} (\bibinfo {year} {2018})}\BibitemShut {NoStop}%
\bibitem [{\citenamefont {Tran}\ \emph {et~al.}(2019)\citenamefont {Tran},
  \citenamefont {Moody}, \citenamefont {Wu}, \citenamefont {Lu}, \citenamefont
  {Choi}, \citenamefont {Kim}, \citenamefont {Rai}, \citenamefont {Sanchez},
  \citenamefont {Quan}, \citenamefont {Singh} \emph
  {et~al.}}]{tran2019evidence}%
  \BibitemOpen
  \bibfield  {author} {\bibinfo {author} {\bibfnamefont {K.}~\bibnamefont
  {Tran}}, \bibinfo {author} {\bibfnamefont {G.}~\bibnamefont {Moody}},
  \bibinfo {author} {\bibfnamefont {F.}~\bibnamefont {Wu}}, \bibinfo {author}
  {\bibfnamefont {X.}~\bibnamefont {Lu}}, \bibinfo {author} {\bibfnamefont
  {J.}~\bibnamefont {Choi}}, \bibinfo {author} {\bibfnamefont {K.}~\bibnamefont
  {Kim}}, \bibinfo {author} {\bibfnamefont {A.}~\bibnamefont {Rai}}, \bibinfo
  {author} {\bibfnamefont {D.~A.}\ \bibnamefont {Sanchez}}, \bibinfo {author}
  {\bibfnamefont {J.}~\bibnamefont {Quan}}, \bibinfo {author} {\bibfnamefont
  {A.}~\bibnamefont {Singh}}, \emph {et~al.},\ }\bibfield  {title} {\bibinfo
  {title} {Evidence for moir{\'e} excitons in van der {W}aals
  heterostructures},\ }\href {https://doi.org/10.1038/s41586-019-0975-z}
  {\bibfield  {journal} {\bibinfo  {journal} {Nature}\ }\textbf {\bibinfo
  {volume} {567}},\ \bibinfo {pages} {71} (\bibinfo {year} {2019})}\BibitemShut
  {NoStop}%
\bibitem [{\citenamefont {Docherty}\ \emph {et~al.}(2014)\citenamefont
  {Docherty}, \citenamefont {Parkinson}, \citenamefont {Joyce}, \citenamefont
  {Chiu}, \citenamefont {Chen}, \citenamefont {Lee}, \citenamefont {Li},
  \citenamefont {Herz},\ and\ \citenamefont
  {Johnston}}]{docherty2014ultrafast}%
  \BibitemOpen
  \bibfield  {author} {\bibinfo {author} {\bibfnamefont {C.~J.}\ \bibnamefont
  {Docherty}}, \bibinfo {author} {\bibfnamefont {P.}~\bibnamefont {Parkinson}},
  \bibinfo {author} {\bibfnamefont {H.~J.}\ \bibnamefont {Joyce}}, \bibinfo
  {author} {\bibfnamefont {M.-H.}\ \bibnamefont {Chiu}}, \bibinfo {author}
  {\bibfnamefont {C.-H.}\ \bibnamefont {Chen}}, \bibinfo {author}
  {\bibfnamefont {M.-Y.}\ \bibnamefont {Lee}}, \bibinfo {author} {\bibfnamefont
  {L.-J.}\ \bibnamefont {Li}}, \bibinfo {author} {\bibfnamefont {L.~M.}\
  \bibnamefont {Herz}},\ and\ \bibinfo {author} {\bibfnamefont {M.~B.}\
  \bibnamefont {Johnston}},\ }\bibfield  {title} {\bibinfo {title} {Ultrafast
  transient terahertz conductivity of monolayer {M}o{S}\textsubscript{2} and
  {W}{S}e\textsubscript{2} grown by chemical vapor deposition},\ }\href
  {https://doi.org/10.1021/nn5034746} {\bibfield  {journal} {\bibinfo
  {journal} {ACS Nano}\ }\textbf {\bibinfo {volume} {8}},\ \bibinfo {pages}
  {11147} (\bibinfo {year} {2014})}\BibitemShut {NoStop}%
\bibitem [{\citenamefont {Mak}\ \emph {et~al.}(2013)\citenamefont {Mak},
  \citenamefont {He}, \citenamefont {Lee}, \citenamefont {Lee}, \citenamefont
  {Hone}, \citenamefont {Heinz},\ and\ \citenamefont {Shan}}]{mak2013tightly}%
  \BibitemOpen
  \bibfield  {author} {\bibinfo {author} {\bibfnamefont {K.~F.}\ \bibnamefont
  {Mak}}, \bibinfo {author} {\bibfnamefont {K.}~\bibnamefont {He}}, \bibinfo
  {author} {\bibfnamefont {C.}~\bibnamefont {Lee}}, \bibinfo {author}
  {\bibfnamefont {G.~H.}\ \bibnamefont {Lee}}, \bibinfo {author} {\bibfnamefont
  {J.}~\bibnamefont {Hone}}, \bibinfo {author} {\bibfnamefont {T.~F.}\
  \bibnamefont {Heinz}},\ and\ \bibinfo {author} {\bibfnamefont
  {J.}~\bibnamefont {Shan}},\ }\bibfield  {title} {\bibinfo {title} {Tightly
  bound trions in monolayer {M}o{S}\textsubscript{2}},\ }\href
  {https://doi.org/10.1038/nmat35052} {\bibfield  {journal} {\bibinfo
  {journal} {Nature Materials}\ }\textbf {\bibinfo {volume} {12}},\ \bibinfo
  {pages} {207} (\bibinfo {year} {2013})}\BibitemShut {NoStop}%
\bibitem [{\citenamefont {Ross}\ \emph {et~al.}(2013)\citenamefont {Ross},
  \citenamefont {Wu}, \citenamefont {Yu}, \citenamefont {Ghimire},
  \citenamefont {Jones}, \citenamefont {Aivazian}, \citenamefont {Yan},
  \citenamefont {Mandrus}, \citenamefont {Xiao}, \citenamefont {Yao} \emph
  {et~al.}}]{ross2013electrical}%
  \BibitemOpen
  \bibfield  {author} {\bibinfo {author} {\bibfnamefont {J.~S.}\ \bibnamefont
  {Ross}}, \bibinfo {author} {\bibfnamefont {S.}~\bibnamefont {Wu}}, \bibinfo
  {author} {\bibfnamefont {H.}~\bibnamefont {Yu}}, \bibinfo {author}
  {\bibfnamefont {N.~J.}\ \bibnamefont {Ghimire}}, \bibinfo {author}
  {\bibfnamefont {A.~M.}\ \bibnamefont {Jones}}, \bibinfo {author}
  {\bibfnamefont {G.}~\bibnamefont {Aivazian}}, \bibinfo {author}
  {\bibfnamefont {J.}~\bibnamefont {Yan}}, \bibinfo {author} {\bibfnamefont
  {D.~G.}\ \bibnamefont {Mandrus}}, \bibinfo {author} {\bibfnamefont
  {D.}~\bibnamefont {Xiao}}, \bibinfo {author} {\bibfnamefont {W.}~\bibnamefont
  {Yao}}, \emph {et~al.},\ }\bibfield  {title} {\bibinfo {title} {Electrical
  control of neutral and charged excitons in a monolayer semiconductor},\
  }\href {https://doi.org/10.1038/ncomms2498} {\bibfield  {journal} {\bibinfo
  {journal} {Nature Communications}\ }\textbf {\bibinfo {volume} {4}},\
  \bibinfo {pages} {1} (\bibinfo {year} {2013})}\BibitemShut {NoStop}%
\bibitem [{\citenamefont {Sidler}\ \emph {et~al.}(2017)\citenamefont {Sidler},
  \citenamefont {Back}, \citenamefont {Cotlet}, \citenamefont {Srivastava},
  \citenamefont {Fink}, \citenamefont {Kroner}, \citenamefont {Demler},\ and\
  \citenamefont {Imamoglu}}]{sidler2017fermi}%
  \BibitemOpen
  \bibfield  {author} {\bibinfo {author} {\bibfnamefont {M.}~\bibnamefont
  {Sidler}}, \bibinfo {author} {\bibfnamefont {P.}~\bibnamefont {Back}},
  \bibinfo {author} {\bibfnamefont {O.}~\bibnamefont {Cotlet}}, \bibinfo
  {author} {\bibfnamefont {A.}~\bibnamefont {Srivastava}}, \bibinfo {author}
  {\bibfnamefont {T.}~\bibnamefont {Fink}}, \bibinfo {author} {\bibfnamefont
  {M.}~\bibnamefont {Kroner}}, \bibinfo {author} {\bibfnamefont
  {E.}~\bibnamefont {Demler}},\ and\ \bibinfo {author} {\bibfnamefont
  {A.}~\bibnamefont {Imamoglu}},\ }\bibfield  {title} {\bibinfo {title} {Fermi
  polaron-polaritons in charge-tunable atomically thin semiconductors},\ }\href
  {https://doi.org/10.1038/nphys3949} {\bibfield  {journal} {\bibinfo
  {journal} {Nature Physics}\ }\textbf {\bibinfo {volume} {13}},\ \bibinfo
  {pages} {255} (\bibinfo {year} {2017})}\BibitemShut {NoStop}%
\bibitem [{\citenamefont {Efimkin}\ and\ \citenamefont
  {MacDonald}(2017)}]{efimkin2017many}%
  \BibitemOpen
  \bibfield  {author} {\bibinfo {author} {\bibfnamefont {D.~K.}\ \bibnamefont
  {Efimkin}}\ and\ \bibinfo {author} {\bibfnamefont {A.~H.}\ \bibnamefont
  {MacDonald}},\ }\bibfield  {title} {\bibinfo {title} {Many-body theory of
  trion absorption features in two-dimensional semiconductors},\ }\href
  {https://doi.org/10.1103/PhysRevB.95.035417} {\bibfield  {journal} {\bibinfo
  {journal} {Physical Review B}\ }\textbf {\bibinfo {volume} {95}},\ \bibinfo
  {pages} {035417} (\bibinfo {year} {2017})}\BibitemShut {NoStop}%
\bibitem [{\citenamefont {Glazov}(2020{\natexlab{a}})}]{glazov2020optical}%
  \BibitemOpen
  \bibfield  {author} {\bibinfo {author} {\bibfnamefont {M.~M.}\ \bibnamefont
  {Glazov}},\ }\bibfield  {title} {\bibinfo {title} {Optical properties of
  charged excitons in two-dimensional semiconductors},\ }\href
  {https://doi.org/10.1063/5.0012475} {\bibfield  {journal} {\bibinfo
  {journal} {The Journal of Chemical Physics}\ }\textbf {\bibinfo {volume}
  {153}},\ \bibinfo {pages} {034703} (\bibinfo {year}
  {2020}{\natexlab{a}})}\BibitemShut {NoStop}%
\bibitem [{\citenamefont {Rana}\ \emph {et~al.}(2020)\citenamefont {Rana},
  \citenamefont {Koksal},\ and\ \citenamefont {Manolatou}}]{rana2020many}%
  \BibitemOpen
  \bibfield  {author} {\bibinfo {author} {\bibfnamefont {F.}~\bibnamefont
  {Rana}}, \bibinfo {author} {\bibfnamefont {O.}~\bibnamefont {Koksal}},\ and\
  \bibinfo {author} {\bibfnamefont {C.}~\bibnamefont {Manolatou}},\ }\bibfield
  {title} {\bibinfo {title} {Many-body theory of the optical conductivity of
  excitons and trions in two-dimensional materials},\ }\href
  {https://doi.org/10.1103/PhysRevB.102.085304} {\bibfield  {journal} {\bibinfo
   {journal} {Physical Review B}\ }\textbf {\bibinfo {volume} {102}},\ \bibinfo
  {pages} {085304} (\bibinfo {year} {2020})}\BibitemShut {NoStop}%
\bibitem [{\citenamefont {Imamoglu}\ \emph {et~al.}(2021)\citenamefont
  {Imamoglu}, \citenamefont {Cotlet},\ and\ \citenamefont
  {Schmidt}}]{imamoglu2021exciton}%
  \BibitemOpen
  \bibfield  {author} {\bibinfo {author} {\bibfnamefont {A.}~\bibnamefont
  {Imamoglu}}, \bibinfo {author} {\bibfnamefont {O.}~\bibnamefont {Cotlet}},\
  and\ \bibinfo {author} {\bibfnamefont {R.}~\bibnamefont {Schmidt}},\
  }\bibfield  {title} {\bibinfo {title} {Exciton--polarons in two-dimensional
  semiconductors and the {T}avis--{C}ummings model},\ }\href
  {https://doi.org/10.5802/crphys.47} {\bibfield  {journal} {\bibinfo
  {journal} {Comptes Rendus. Physique}\ }\textbf {\bibinfo {volume} {22}},\
  \bibinfo {pages} {1} (\bibinfo {year} {2021})}\BibitemShut {NoStop}%
\bibitem [{\citenamefont {Efimkin}\ \emph {et~al.}(2021)\citenamefont
  {Efimkin}, \citenamefont {Laird}, \citenamefont {Levinsen}, \citenamefont
  {Parish},\ and\ \citenamefont {MacDonald}}]{efimkin2021electron}%
  \BibitemOpen
  \bibfield  {author} {\bibinfo {author} {\bibfnamefont {D.~K.}\ \bibnamefont
  {Efimkin}}, \bibinfo {author} {\bibfnamefont {E.~K.}\ \bibnamefont {Laird}},
  \bibinfo {author} {\bibfnamefont {J.}~\bibnamefont {Levinsen}}, \bibinfo
  {author} {\bibfnamefont {M.~M.}\ \bibnamefont {Parish}},\ and\ \bibinfo
  {author} {\bibfnamefont {A.~H.}\ \bibnamefont {MacDonald}},\ }\bibfield
  {title} {\bibinfo {title} {Electron-exciton interactions in the
  exciton-polaron problem},\ }\href
  {https://doi.org/10.1103/PhysRevB.103.075417} {\bibfield  {journal} {\bibinfo
   {journal} {Physical Review B}\ }\textbf {\bibinfo {volume} {103}},\ \bibinfo
  {pages} {075417} (\bibinfo {year} {2021})}\BibitemShut {NoStop}%
\bibitem [{\citenamefont {Katsch}\ and\ \citenamefont
  {Knorr}(2022)}]{katsch2022excitonic}%
  \BibitemOpen
  \bibfield  {author} {\bibinfo {author} {\bibfnamefont {F.}~\bibnamefont
  {Katsch}}\ and\ \bibinfo {author} {\bibfnamefont {A.}~\bibnamefont {Knorr}},\
  }\bibfield  {title} {\bibinfo {title} {Excitonic theory of doping-dependent
  optical response in atomically thin semiconductors},\ }\href
  {https://doi.org/10.1103/PhysRevB.105.045301} {\bibfield  {journal} {\bibinfo
   {journal} {Physical Review B}\ }\textbf {\bibinfo {volume} {105}},\ \bibinfo
  {pages} {045301} (\bibinfo {year} {2022})}\BibitemShut {NoStop}%
\bibitem [{\citenamefont {Plechinger}\ \emph {et~al.}(2016)\citenamefont
  {Plechinger}, \citenamefont {Nagler}, \citenamefont {Arora}, \citenamefont
  {Schmidt}, \citenamefont {Chernikov}, \citenamefont {Del~{\'A}guila},
  \citenamefont {Christianen}, \citenamefont {Bratschitsch}, \citenamefont
  {Sch{\"u}ller},\ and\ \citenamefont {Korn}}]{plechinger2016trion}%
  \BibitemOpen
  \bibfield  {author} {\bibinfo {author} {\bibfnamefont {G.}~\bibnamefont
  {Plechinger}}, \bibinfo {author} {\bibfnamefont {P.}~\bibnamefont {Nagler}},
  \bibinfo {author} {\bibfnamefont {A.}~\bibnamefont {Arora}}, \bibinfo
  {author} {\bibfnamefont {R.}~\bibnamefont {Schmidt}}, \bibinfo {author}
  {\bibfnamefont {A.}~\bibnamefont {Chernikov}}, \bibinfo {author}
  {\bibfnamefont {A.~G.}\ \bibnamefont {Del~{\'A}guila}}, \bibinfo {author}
  {\bibfnamefont {P.}~\bibnamefont {Christianen}}, \bibinfo {author}
  {\bibfnamefont {R.}~\bibnamefont {Bratschitsch}}, \bibinfo {author}
  {\bibfnamefont {C.}~\bibnamefont {Sch{\"u}ller}},\ and\ \bibinfo {author}
  {\bibfnamefont {T.}~\bibnamefont {Korn}},\ }\bibfield  {title} {\bibinfo
  {title} {Trion fine structure and coupled spin--valley dynamics in monolayer
  tungsten disulfide},\ }\href {https://doi.org/10.1038/ncomms12715} {\bibfield
   {journal} {\bibinfo  {journal} {Nature Communications}\ }\textbf {\bibinfo
  {volume} {7}},\ \bibinfo {pages} {1} (\bibinfo {year} {2016})}\BibitemShut
  {NoStop}%
\bibitem [{\citenamefont {Courtade}\ \emph {et~al.}(2017)\citenamefont
  {Courtade}, \citenamefont {Semina}, \citenamefont {Manca}, \citenamefont
  {Glazov}, \citenamefont {Robert}, \citenamefont {Cadiz}, \citenamefont
  {Wang}, \citenamefont {Taniguchi}, \citenamefont {Watanabe}, \citenamefont
  {Pierre} \emph {et~al.}}]{courtade2017charged}%
  \BibitemOpen
  \bibfield  {author} {\bibinfo {author} {\bibfnamefont {E.}~\bibnamefont
  {Courtade}}, \bibinfo {author} {\bibfnamefont {M.}~\bibnamefont {Semina}},
  \bibinfo {author} {\bibfnamefont {M.}~\bibnamefont {Manca}}, \bibinfo
  {author} {\bibfnamefont {M.}~\bibnamefont {Glazov}}, \bibinfo {author}
  {\bibfnamefont {C.}~\bibnamefont {Robert}}, \bibinfo {author} {\bibfnamefont
  {F.}~\bibnamefont {Cadiz}}, \bibinfo {author} {\bibfnamefont
  {G.}~\bibnamefont {Wang}}, \bibinfo {author} {\bibfnamefont {T.}~\bibnamefont
  {Taniguchi}}, \bibinfo {author} {\bibfnamefont {K.}~\bibnamefont {Watanabe}},
  \bibinfo {author} {\bibfnamefont {M.}~\bibnamefont {Pierre}}, \emph
  {et~al.},\ }\bibfield  {title} {\bibinfo {title} {Charged excitons in
  monolayer {W}{S}e\textsubscript{2}: Experiment and theory},\ }\href
  {https://doi.org/10.1103/PhysRevB.96.085302} {\bibfield  {journal} {\bibinfo
  {journal} {Physical Review B}\ }\textbf {\bibinfo {volume} {96}},\ \bibinfo
  {pages} {085302} (\bibinfo {year} {2017})}\BibitemShut {NoStop}%
\bibitem [{\citenamefont {Arora}\ \emph {et~al.}(2019)\citenamefont {Arora},
  \citenamefont {Deilmann}, \citenamefont {Reichenauer}, \citenamefont {Kern},
  \citenamefont {de~Vasconcellos}, \citenamefont {Rohlfing},\ and\
  \citenamefont {Bratschitsch}}]{arora2019excited}%
  \BibitemOpen
  \bibfield  {author} {\bibinfo {author} {\bibfnamefont {A.}~\bibnamefont
  {Arora}}, \bibinfo {author} {\bibfnamefont {T.}~\bibnamefont {Deilmann}},
  \bibinfo {author} {\bibfnamefont {T.}~\bibnamefont {Reichenauer}}, \bibinfo
  {author} {\bibfnamefont {J.}~\bibnamefont {Kern}}, \bibinfo {author}
  {\bibfnamefont {S.~M.}\ \bibnamefont {de~Vasconcellos}}, \bibinfo {author}
  {\bibfnamefont {M.}~\bibnamefont {Rohlfing}},\ and\ \bibinfo {author}
  {\bibfnamefont {R.}~\bibnamefont {Bratschitsch}},\ }\bibfield  {title}
  {\bibinfo {title} {Excited-state trions in monolayer
  {W}{S}\textsubscript{2}},\ }\href
  {https://doi.org/10.1103/PhysRevLett.123.167401} {\bibfield  {journal}
  {\bibinfo  {journal} {Physical Review Letters}\ }\textbf {\bibinfo {volume}
  {123}},\ \bibinfo {pages} {167401} (\bibinfo {year} {2019})}\BibitemShut
  {NoStop}%
\bibitem [{\citenamefont {Arora}\ \emph {et~al.}(2020)\citenamefont {Arora},
  \citenamefont {Wessling}, \citenamefont {Deilmann}, \citenamefont
  {Reichenauer}, \citenamefont {Steeger}, \citenamefont {Kossacki},
  \citenamefont {Potemski}, \citenamefont {de~Vasconcellos}, \citenamefont
  {Rohlfing},\ and\ \citenamefont {Bratschitsch}}]{arora2020dark}%
  \BibitemOpen
  \bibfield  {author} {\bibinfo {author} {\bibfnamefont {A.}~\bibnamefont
  {Arora}}, \bibinfo {author} {\bibfnamefont {N.~K.}\ \bibnamefont {Wessling}},
  \bibinfo {author} {\bibfnamefont {T.}~\bibnamefont {Deilmann}}, \bibinfo
  {author} {\bibfnamefont {T.}~\bibnamefont {Reichenauer}}, \bibinfo {author}
  {\bibfnamefont {P.}~\bibnamefont {Steeger}}, \bibinfo {author} {\bibfnamefont
  {P.}~\bibnamefont {Kossacki}}, \bibinfo {author} {\bibfnamefont
  {M.}~\bibnamefont {Potemski}}, \bibinfo {author} {\bibfnamefont {S.~M.}\
  \bibnamefont {de~Vasconcellos}}, \bibinfo {author} {\bibfnamefont
  {M.}~\bibnamefont {Rohlfing}},\ and\ \bibinfo {author} {\bibfnamefont
  {R.}~\bibnamefont {Bratschitsch}},\ }\bibfield  {title} {\bibinfo {title}
  {Dark trions govern the temperature-dependent optical absorption and emission
  of doped atomically thin semiconductors},\ }\href
  {https://doi.org/10.1103/PhysRevB.101.241413} {\bibfield  {journal} {\bibinfo
   {journal} {Physical Review B}\ }\textbf {\bibinfo {volume} {101}},\ \bibinfo
  {pages} {241413} (\bibinfo {year} {2020})}\BibitemShut {NoStop}%
\bibitem [{\citenamefont {Wagner}\ \emph {et~al.}(2020)\citenamefont {Wagner},
  \citenamefont {Wietek}, \citenamefont {Ziegler}, \citenamefont {Semina},
  \citenamefont {Taniguchi}, \citenamefont {Watanabe}, \citenamefont {Zipfel},
  \citenamefont {Glazov},\ and\ \citenamefont
  {Chernikov}}]{wagner2020autoionization}%
  \BibitemOpen
  \bibfield  {author} {\bibinfo {author} {\bibfnamefont {K.}~\bibnamefont
  {Wagner}}, \bibinfo {author} {\bibfnamefont {E.}~\bibnamefont {Wietek}},
  \bibinfo {author} {\bibfnamefont {J.~D.}\ \bibnamefont {Ziegler}}, \bibinfo
  {author} {\bibfnamefont {M.~A.}\ \bibnamefont {Semina}}, \bibinfo {author}
  {\bibfnamefont {T.}~\bibnamefont {Taniguchi}}, \bibinfo {author}
  {\bibfnamefont {K.}~\bibnamefont {Watanabe}}, \bibinfo {author}
  {\bibfnamefont {J.}~\bibnamefont {Zipfel}}, \bibinfo {author} {\bibfnamefont
  {M.~M.}\ \bibnamefont {Glazov}},\ and\ \bibinfo {author} {\bibfnamefont
  {A.}~\bibnamefont {Chernikov}},\ }\bibfield  {title} {\bibinfo {title}
  {Autoionization and dressing of excited excitons by free carriers in
  monolayer {W}{S}e\textsubscript{2}},\ }\href
  {https://doi.org/10.1103/PhysRevLett.125.267401} {\bibfield  {journal}
  {\bibinfo  {journal} {Physical review letters}\ }\textbf {\bibinfo {volume}
  {125}},\ \bibinfo {pages} {267401} (\bibinfo {year} {2020})}\BibitemShut
  {NoStop}%
\bibitem [{\citenamefont {Goldstein}\ \emph {et~al.}(2020)\citenamefont
  {Goldstein}, \citenamefont {Wu}, \citenamefont {Chen}, \citenamefont
  {Taniguchi}, \citenamefont {Watanabe}, \citenamefont {Varga},\ and\
  \citenamefont {Yan}}]{goldstein2020ground}%
  \BibitemOpen
  \bibfield  {author} {\bibinfo {author} {\bibfnamefont {T.}~\bibnamefont
  {Goldstein}}, \bibinfo {author} {\bibfnamefont {Y.-C.}\ \bibnamefont {Wu}},
  \bibinfo {author} {\bibfnamefont {S.-Y.}\ \bibnamefont {Chen}}, \bibinfo
  {author} {\bibfnamefont {T.}~\bibnamefont {Taniguchi}}, \bibinfo {author}
  {\bibfnamefont {K.}~\bibnamefont {Watanabe}}, \bibinfo {author}
  {\bibfnamefont {K.}~\bibnamefont {Varga}},\ and\ \bibinfo {author}
  {\bibfnamefont {J.}~\bibnamefont {Yan}},\ }\bibfield  {title} {\bibinfo
  {title} {Ground and excited state exciton polarons in monolayer
  {M}o{S}e\textsubscript{2}},\ }\href {https://doi.org/10.1063/5.0013092}
  {\bibfield  {journal} {\bibinfo  {journal} {The Journal of Chemical Physics}\
  }\textbf {\bibinfo {volume} {153}},\ \bibinfo {pages} {071101} (\bibinfo
  {year} {2020})}\BibitemShut {NoStop}%
\bibitem [{\citenamefont {He}\ \emph {et~al.}(2020)\citenamefont {He},
  \citenamefont {Rivera}, \citenamefont {Van~Tuan}, \citenamefont {Wilson},
  \citenamefont {Yang}, \citenamefont {Taniguchi}, \citenamefont {Watanabe},
  \citenamefont {Yan}, \citenamefont {Mandrus}, \citenamefont {Yu} \emph
  {et~al.}}]{he2020valley}%
  \BibitemOpen
  \bibfield  {author} {\bibinfo {author} {\bibfnamefont {M.}~\bibnamefont
  {He}}, \bibinfo {author} {\bibfnamefont {P.}~\bibnamefont {Rivera}}, \bibinfo
  {author} {\bibfnamefont {D.}~\bibnamefont {Van~Tuan}}, \bibinfo {author}
  {\bibfnamefont {N.~P.}\ \bibnamefont {Wilson}}, \bibinfo {author}
  {\bibfnamefont {M.}~\bibnamefont {Yang}}, \bibinfo {author} {\bibfnamefont
  {T.}~\bibnamefont {Taniguchi}}, \bibinfo {author} {\bibfnamefont
  {K.}~\bibnamefont {Watanabe}}, \bibinfo {author} {\bibfnamefont
  {J.}~\bibnamefont {Yan}}, \bibinfo {author} {\bibfnamefont {D.~G.}\
  \bibnamefont {Mandrus}}, \bibinfo {author} {\bibfnamefont {H.}~\bibnamefont
  {Yu}}, \emph {et~al.},\ }\bibfield  {title} {\bibinfo {title} {Valley phonons
  and exciton complexes in a monolayer semiconductor},\ }\href
  {https://doi.org/10.1038/s41467-020-14472-0} {\bibfield  {journal} {\bibinfo
  {journal} {Nature Communications}\ }\textbf {\bibinfo {volume} {11}},\
  \bibinfo {pages} {1} (\bibinfo {year} {2020})}\BibitemShut {NoStop}%
\bibitem [{\citenamefont {Liu}\ \emph {et~al.}(2021)\citenamefont {Liu},
  \citenamefont {van Baren}, \citenamefont {Lu}, \citenamefont {Taniguchi},
  \citenamefont {Watanabe}, \citenamefont {Smirnov}, \citenamefont {Chang},\
  and\ \citenamefont {Lui}}]{liu2021exciton}%
  \BibitemOpen
  \bibfield  {author} {\bibinfo {author} {\bibfnamefont {E.}~\bibnamefont
  {Liu}}, \bibinfo {author} {\bibfnamefont {J.}~\bibnamefont {van Baren}},
  \bibinfo {author} {\bibfnamefont {Z.}~\bibnamefont {Lu}}, \bibinfo {author}
  {\bibfnamefont {T.}~\bibnamefont {Taniguchi}}, \bibinfo {author}
  {\bibfnamefont {K.}~\bibnamefont {Watanabe}}, \bibinfo {author}
  {\bibfnamefont {D.}~\bibnamefont {Smirnov}}, \bibinfo {author} {\bibfnamefont
  {Y.-C.}\ \bibnamefont {Chang}},\ and\ \bibinfo {author} {\bibfnamefont
  {C.~H.}\ \bibnamefont {Lui}},\ }\bibfield  {title} {\bibinfo {title}
  {Exciton-polaron rydberg states in monolayer {M}o{S}e\textsubscript{2} and
  {W}{S}e\textsubscript{2}},\ }\href
  {https://doi.org/10.1038/s41467-021-26304-w} {\bibfield  {journal} {\bibinfo
  {journal} {Nature Communications}\ }\textbf {\bibinfo {volume} {12}},\
  \bibinfo {pages} {1} (\bibinfo {year} {2021})}\BibitemShut {NoStop}%
\bibitem [{\citenamefont {Yang}\ \emph {et~al.}(2022)\citenamefont {Yang},
  \citenamefont {Ren}, \citenamefont {Robert}, \citenamefont {Van~Tuan},
  \citenamefont {Lombez}, \citenamefont {Urbaszek}, \citenamefont {Marie},\
  and\ \citenamefont {Dery}}]{yang2022relaxation}%
  \BibitemOpen
  \bibfield  {author} {\bibinfo {author} {\bibfnamefont {M.}~\bibnamefont
  {Yang}}, \bibinfo {author} {\bibfnamefont {L.}~\bibnamefont {Ren}}, \bibinfo
  {author} {\bibfnamefont {C.}~\bibnamefont {Robert}}, \bibinfo {author}
  {\bibfnamefont {D.}~\bibnamefont {Van~Tuan}}, \bibinfo {author}
  {\bibfnamefont {L.}~\bibnamefont {Lombez}}, \bibinfo {author} {\bibfnamefont
  {B.}~\bibnamefont {Urbaszek}}, \bibinfo {author} {\bibfnamefont
  {X.}~\bibnamefont {Marie}},\ and\ \bibinfo {author} {\bibfnamefont
  {H.}~\bibnamefont {Dery}},\ }\bibfield  {title} {\bibinfo {title} {Relaxation
  and darkening of excitonic complexes in electrostatically doped monolayer
  {W}{S}e\textsubscript{2}: Roles of exciton-electron and trion-electron
  interactions},\ }\href {https://doi.org/10.1103/PhysRevB.105.085302}
  {\bibfield  {journal} {\bibinfo  {journal} {Physical Review B}\ }\textbf
  {\bibinfo {volume} {105}},\ \bibinfo {pages} {085302} (\bibinfo {year}
  {2022})}\BibitemShut {NoStop}%
\bibitem [{\citenamefont {Klein}\ \emph {et~al.}(2022)\citenamefont {Klein},
  \citenamefont {Florian}, \citenamefont {H{\"o}tger}, \citenamefont
  {Steinhoff}, \citenamefont {Delhomme}, \citenamefont {Taniguchi},
  \citenamefont {Watanabe}, \citenamefont {Jahnke}, \citenamefont {Holleitner},
  \citenamefont {Potemski} \emph {et~al.}}]{klein2022trions}%
  \BibitemOpen
  \bibfield  {author} {\bibinfo {author} {\bibfnamefont {J.}~\bibnamefont
  {Klein}}, \bibinfo {author} {\bibfnamefont {M.}~\bibnamefont {Florian}},
  \bibinfo {author} {\bibfnamefont {A.}~\bibnamefont {H{\"o}tger}}, \bibinfo
  {author} {\bibfnamefont {A.}~\bibnamefont {Steinhoff}}, \bibinfo {author}
  {\bibfnamefont {A.}~\bibnamefont {Delhomme}}, \bibinfo {author}
  {\bibfnamefont {T.}~\bibnamefont {Taniguchi}}, \bibinfo {author}
  {\bibfnamefont {K.}~\bibnamefont {Watanabe}}, \bibinfo {author}
  {\bibfnamefont {F.}~\bibnamefont {Jahnke}}, \bibinfo {author} {\bibfnamefont
  {A.~W.}\ \bibnamefont {Holleitner}}, \bibinfo {author} {\bibfnamefont
  {M.}~\bibnamefont {Potemski}}, \emph {et~al.},\ }\bibfield  {title} {\bibinfo
  {title} {Trions in {M}o{S}\textsubscript{2} are quantum superpositions of
  intra-and intervalley spin states},\ }\href
  {https://doi.org/10.1103/PhysRevB.105.L041302} {\bibfield  {journal}
  {\bibinfo  {journal} {Physical Review B}\ }\textbf {\bibinfo {volume}
  {105}},\ \bibinfo {pages} {L041302} (\bibinfo {year} {2022})}\BibitemShut
  {NoStop}%
\bibitem [{\citenamefont {Deilmann}\ and\ \citenamefont
  {Thygesen}(2017)}]{deilmann2017dark}%
  \BibitemOpen
  \bibfield  {author} {\bibinfo {author} {\bibfnamefont {T.}~\bibnamefont
  {Deilmann}}\ and\ \bibinfo {author} {\bibfnamefont {K.~S.}\ \bibnamefont
  {Thygesen}},\ }\bibfield  {title} {\bibinfo {title} {Dark excitations in
  monolayer transition metal dichalcogenides},\ }\href
  {https://doi.org/10.1103/PhysRevB.96.201113} {\bibfield  {journal} {\bibinfo
  {journal} {Physical Review B}\ }\textbf {\bibinfo {volume} {96}},\ \bibinfo
  {pages} {201113} (\bibinfo {year} {2017})}\BibitemShut {NoStop}%
\bibitem [{\citenamefont {Fey}\ \emph {et~al.}(2020)\citenamefont {Fey},
  \citenamefont {Schmelcher}, \citenamefont {Imamoglu},\ and\ \citenamefont
  {Schmidt}}]{fey2020theory}%
  \BibitemOpen
  \bibfield  {author} {\bibinfo {author} {\bibfnamefont {C.}~\bibnamefont
  {Fey}}, \bibinfo {author} {\bibfnamefont {P.}~\bibnamefont {Schmelcher}},
  \bibinfo {author} {\bibfnamefont {A.}~\bibnamefont {Imamoglu}},\ and\
  \bibinfo {author} {\bibfnamefont {R.}~\bibnamefont {Schmidt}},\ }\bibfield
  {title} {\bibinfo {title} {Theory of exciton-electron scattering in
  atomically thin semiconductors},\ }\href
  {https://doi.org/10.1103/PhysRevB.101.195417} {\bibfield  {journal} {\bibinfo
   {journal} {Physical Review B}\ }\textbf {\bibinfo {volume} {101}},\ \bibinfo
  {pages} {195417} (\bibinfo {year} {2020})}\BibitemShut {NoStop}%
\bibitem [{\citenamefont {Singh}\ \emph
  {et~al.}(2016{\natexlab{a}})\citenamefont {Singh}, \citenamefont {Moody},
  \citenamefont {Tran}, \citenamefont {Scott}, \citenamefont {Overbeck},
  \citenamefont {Bergh{\"a}user}, \citenamefont {Schaibley}, \citenamefont
  {Seifert}, \citenamefont {Pleskot}, \citenamefont {Gabor} \emph
  {et~al.}}]{singh2016trion}%
  \BibitemOpen
  \bibfield  {author} {\bibinfo {author} {\bibfnamefont {A.}~\bibnamefont
  {Singh}}, \bibinfo {author} {\bibfnamefont {G.}~\bibnamefont {Moody}},
  \bibinfo {author} {\bibfnamefont {K.}~\bibnamefont {Tran}}, \bibinfo {author}
  {\bibfnamefont {M.~E.}\ \bibnamefont {Scott}}, \bibinfo {author}
  {\bibfnamefont {V.}~\bibnamefont {Overbeck}}, \bibinfo {author}
  {\bibfnamefont {G.}~\bibnamefont {Bergh{\"a}user}}, \bibinfo {author}
  {\bibfnamefont {J.}~\bibnamefont {Schaibley}}, \bibinfo {author}
  {\bibfnamefont {E.~J.}\ \bibnamefont {Seifert}}, \bibinfo {author}
  {\bibfnamefont {D.}~\bibnamefont {Pleskot}}, \bibinfo {author} {\bibfnamefont
  {N.~M.}\ \bibnamefont {Gabor}}, \emph {et~al.},\ }\bibfield  {title}
  {\bibinfo {title} {Trion formation dynamics in monolayer transition metal
  dichalcogenides},\ }\href {https://doi.org/10.1103/PhysRevB.93.041401}
  {\bibfield  {journal} {\bibinfo  {journal} {Physical Review B}\ }\textbf
  {\bibinfo {volume} {93}},\ \bibinfo {pages} {041401} (\bibinfo {year}
  {2016}{\natexlab{a}})}\BibitemShut {NoStop}%
\bibitem [{\citenamefont {Wang}\ \emph {et~al.}(2016)\citenamefont {Wang},
  \citenamefont {Zhang}, \citenamefont {Chan}, \citenamefont {Manolatou},
  \citenamefont {Tiwari},\ and\ \citenamefont {Rana}}]{wang2016radiative}%
  \BibitemOpen
  \bibfield  {author} {\bibinfo {author} {\bibfnamefont {H.}~\bibnamefont
  {Wang}}, \bibinfo {author} {\bibfnamefont {C.}~\bibnamefont {Zhang}},
  \bibinfo {author} {\bibfnamefont {W.}~\bibnamefont {Chan}}, \bibinfo {author}
  {\bibfnamefont {C.}~\bibnamefont {Manolatou}}, \bibinfo {author}
  {\bibfnamefont {S.}~\bibnamefont {Tiwari}},\ and\ \bibinfo {author}
  {\bibfnamefont {F.}~\bibnamefont {Rana}},\ }\bibfield  {title} {\bibinfo
  {title} {Radiative lifetimes of excitons and trions in monolayers of the
  metal dichalcogenide {M}o{S}\textsubscript{2}},\ }\href
  {https://doi.org/10.1103/PhysRevB.93.045407} {\bibfield  {journal} {\bibinfo
  {journal} {Physical Review B}\ }\textbf {\bibinfo {volume} {93}},\ \bibinfo
  {pages} {045407} (\bibinfo {year} {2016})}\BibitemShut {NoStop}%
\bibitem [{\citenamefont {Singh}\ \emph
  {et~al.}(2016{\natexlab{b}})\citenamefont {Singh}, \citenamefont {Tran},
  \citenamefont {Kolarczik}, \citenamefont {Seifert}, \citenamefont {Wang},
  \citenamefont {Hao}, \citenamefont {Pleskot}, \citenamefont {Gabor},
  \citenamefont {Helmrich}, \citenamefont {Owschimikow} \emph
  {et~al.}}]{singh2016long}%
  \BibitemOpen
  \bibfield  {author} {\bibinfo {author} {\bibfnamefont {A.}~\bibnamefont
  {Singh}}, \bibinfo {author} {\bibfnamefont {K.}~\bibnamefont {Tran}},
  \bibinfo {author} {\bibfnamefont {M.}~\bibnamefont {Kolarczik}}, \bibinfo
  {author} {\bibfnamefont {J.}~\bibnamefont {Seifert}}, \bibinfo {author}
  {\bibfnamefont {Y.}~\bibnamefont {Wang}}, \bibinfo {author} {\bibfnamefont
  {K.}~\bibnamefont {Hao}}, \bibinfo {author} {\bibfnamefont {D.}~\bibnamefont
  {Pleskot}}, \bibinfo {author} {\bibfnamefont {N.~M.}\ \bibnamefont {Gabor}},
  \bibinfo {author} {\bibfnamefont {S.}~\bibnamefont {Helmrich}}, \bibinfo
  {author} {\bibfnamefont {N.}~\bibnamefont {Owschimikow}}, \emph {et~al.},\
  }\bibfield  {title} {\bibinfo {title} {Long-lived valley polarization of
  intravalley trions in monolayer {W}{S}e\textsubscript{2}},\ }\href
  {https://doi.org/10.1103/PhysRevLett.117.257402} {\bibfield  {journal}
  {\bibinfo  {journal} {Physical Review Letters}\ }\textbf {\bibinfo {volume}
  {117}},\ \bibinfo {pages} {257402} (\bibinfo {year}
  {2016}{\natexlab{b}})}\BibitemShut {NoStop}%
\bibitem [{\citenamefont {Uddin}\ \emph {et~al.}(2020)\citenamefont {Uddin},
  \citenamefont {Kim}, \citenamefont {Lorenzon}, \citenamefont {Yeh},
  \citenamefont {Lien}, \citenamefont {Barnard}, \citenamefont {Htoon},
  \citenamefont {Weber-Bargioni},\ and\ \citenamefont
  {Javey}}]{uddin2020neutral}%
  \BibitemOpen
  \bibfield  {author} {\bibinfo {author} {\bibfnamefont {S.~Z.}\ \bibnamefont
  {Uddin}}, \bibinfo {author} {\bibfnamefont {H.}~\bibnamefont {Kim}}, \bibinfo
  {author} {\bibfnamefont {M.}~\bibnamefont {Lorenzon}}, \bibinfo {author}
  {\bibfnamefont {M.}~\bibnamefont {Yeh}}, \bibinfo {author} {\bibfnamefont
  {D.-H.}\ \bibnamefont {Lien}}, \bibinfo {author} {\bibfnamefont {E.~S.}\
  \bibnamefont {Barnard}}, \bibinfo {author} {\bibfnamefont {H.}~\bibnamefont
  {Htoon}}, \bibinfo {author} {\bibfnamefont {A.}~\bibnamefont
  {Weber-Bargioni}},\ and\ \bibinfo {author} {\bibfnamefont {A.}~\bibnamefont
  {Javey}},\ }\bibfield  {title} {\bibinfo {title} {Neutral exciton diffusion
  in monolayer {M}o{S}\textsubscript{2}},\ }\href
  {https://doi.org/10.1021/acsnano.0c05305} {\bibfield  {journal} {\bibinfo
  {journal} {ACS nano}\ }\textbf {\bibinfo {volume} {14}},\ \bibinfo {pages}
  {13433} (\bibinfo {year} {2020})}\BibitemShut {NoStop}%
\bibitem [{\citenamefont {Kim}\ \emph {et~al.}(2021)\citenamefont {Kim},
  \citenamefont {Luo}, \citenamefont {Rhodes}, \citenamefont {Bai},
  \citenamefont {Wang}, \citenamefont {Liu}, \citenamefont {Jordan},
  \citenamefont {Huang}, \citenamefont {Li}, \citenamefont {Taniguchi} \emph
  {et~al.}}]{kim2021free}%
  \BibitemOpen
  \bibfield  {author} {\bibinfo {author} {\bibfnamefont {B.}~\bibnamefont
  {Kim}}, \bibinfo {author} {\bibfnamefont {Y.}~\bibnamefont {Luo}}, \bibinfo
  {author} {\bibfnamefont {D.}~\bibnamefont {Rhodes}}, \bibinfo {author}
  {\bibfnamefont {Y.}~\bibnamefont {Bai}}, \bibinfo {author} {\bibfnamefont
  {J.}~\bibnamefont {Wang}}, \bibinfo {author} {\bibfnamefont {S.}~\bibnamefont
  {Liu}}, \bibinfo {author} {\bibfnamefont {A.}~\bibnamefont {Jordan}},
  \bibinfo {author} {\bibfnamefont {B.}~\bibnamefont {Huang}}, \bibinfo
  {author} {\bibfnamefont {Z.}~\bibnamefont {Li}}, \bibinfo {author}
  {\bibfnamefont {T.}~\bibnamefont {Taniguchi}}, \emph {et~al.},\ }\bibfield
  {title} {\bibinfo {title} {Free trions with near-unity quantum yield in
  monolayer {M}o{S}e\textsubscript{2}},\ }\href
  {https://doi.org/10.1021/acsnano.1c04331} {\bibfield  {journal} {\bibinfo
  {journal} {ACS nano}\ } (\bibinfo {year} {2021})}\BibitemShut {NoStop}%
\bibitem [{\citenamefont {Kato}\ and\ \citenamefont
  {Kaneko}(2016)}]{kato2016transport}%
  \BibitemOpen
  \bibfield  {author} {\bibinfo {author} {\bibfnamefont {T.}~\bibnamefont
  {Kato}}\ and\ \bibinfo {author} {\bibfnamefont {T.}~\bibnamefont {Kaneko}},\
  }\bibfield  {title} {\bibinfo {title} {Transport dynamics of neutral excitons
  and trions in monolayer {W}{S}\textsubscript{2}},\ }\href
  {https://doi.org/10.1021/acsnano.6b05580} {\bibfield  {journal} {\bibinfo
  {journal} {ACS nano}\ }\textbf {\bibinfo {volume} {10}},\ \bibinfo {pages}
  {9687} (\bibinfo {year} {2016})}\BibitemShut {NoStop}%
\bibitem [{\citenamefont {Cadiz}\ \emph {et~al.}(2018)\citenamefont {Cadiz},
  \citenamefont {Robert}, \citenamefont {Courtade}, \citenamefont {Manca},
  \citenamefont {Martinelli}, \citenamefont {Taniguchi}, \citenamefont
  {Watanabe}, \citenamefont {Amand}, \citenamefont {Rowe}, \citenamefont
  {Paget} \emph {et~al.}}]{cadiz2018exciton}%
  \BibitemOpen
  \bibfield  {author} {\bibinfo {author} {\bibfnamefont {F.}~\bibnamefont
  {Cadiz}}, \bibinfo {author} {\bibfnamefont {C.}~\bibnamefont {Robert}},
  \bibinfo {author} {\bibfnamefont {E.}~\bibnamefont {Courtade}}, \bibinfo
  {author} {\bibfnamefont {M.}~\bibnamefont {Manca}}, \bibinfo {author}
  {\bibfnamefont {L.}~\bibnamefont {Martinelli}}, \bibinfo {author}
  {\bibfnamefont {T.}~\bibnamefont {Taniguchi}}, \bibinfo {author}
  {\bibfnamefont {K.}~\bibnamefont {Watanabe}}, \bibinfo {author}
  {\bibfnamefont {T.}~\bibnamefont {Amand}}, \bibinfo {author} {\bibfnamefont
  {A.}~\bibnamefont {Rowe}}, \bibinfo {author} {\bibfnamefont {D.}~\bibnamefont
  {Paget}}, \emph {et~al.},\ }\bibfield  {title} {\bibinfo {title} {Exciton
  diffusion in {W}{S}e\textsubscript{2} monolayers embedded in a van der
  {W}aals heterostructure},\ }\href {https://doi.org/10.1063/1.5026478}
  {\bibfield  {journal} {\bibinfo  {journal} {Applied Physics Letters}\
  }\textbf {\bibinfo {volume} {112}},\ \bibinfo {pages} {152106} (\bibinfo
  {year} {2018})}\BibitemShut {NoStop}%
\bibitem [{\citenamefont {Park}\ \emph {et~al.}(2021)\citenamefont {Park},
  \citenamefont {Han}, \citenamefont {Boule}, \citenamefont {Paget},
  \citenamefont {Rowe}, \citenamefont {Sirotti}, \citenamefont {Taniguchi},
  \citenamefont {Watanabe}, \citenamefont {Robert}, \citenamefont {Lombez}
  \emph {et~al.}}]{park2021imaging}%
  \BibitemOpen
  \bibfield  {author} {\bibinfo {author} {\bibfnamefont {S.}~\bibnamefont
  {Park}}, \bibinfo {author} {\bibfnamefont {B.}~\bibnamefont {Han}}, \bibinfo
  {author} {\bibfnamefont {C.}~\bibnamefont {Boule}}, \bibinfo {author}
  {\bibfnamefont {D.}~\bibnamefont {Paget}}, \bibinfo {author} {\bibfnamefont
  {A.~C.}\ \bibnamefont {Rowe}}, \bibinfo {author} {\bibfnamefont
  {F.}~\bibnamefont {Sirotti}}, \bibinfo {author} {\bibfnamefont
  {T.}~\bibnamefont {Taniguchi}}, \bibinfo {author} {\bibfnamefont
  {K.}~\bibnamefont {Watanabe}}, \bibinfo {author} {\bibfnamefont
  {C.}~\bibnamefont {Robert}}, \bibinfo {author} {\bibfnamefont
  {L.}~\bibnamefont {Lombez}}, \emph {et~al.},\ }\bibfield  {title} {\bibinfo
  {title} {Imaging seebeck drift of excitons and trions in
  {M}o{S}e\textsubscript{2} monolayers},\ }\href
  {https://doi.org/10.1088/2053-1583/ac171f} {\bibfield  {journal} {\bibinfo
  {journal} {2D Materials}\ }\textbf {\bibinfo {volume} {8}},\ \bibinfo {pages}
  {045014} (\bibinfo {year} {2021})}\BibitemShut {NoStop}%
\bibitem [{\citenamefont {Cheng}\ \emph {et~al.}(2021)\citenamefont {Cheng},
  \citenamefont {Li}, \citenamefont {Jin}, \citenamefont {Zhang},\ and\
  \citenamefont {Wang}}]{cheng2021observation}%
  \BibitemOpen
  \bibfield  {author} {\bibinfo {author} {\bibfnamefont {G.}~\bibnamefont
  {Cheng}}, \bibinfo {author} {\bibfnamefont {B.}~\bibnamefont {Li}}, \bibinfo
  {author} {\bibfnamefont {Z.}~\bibnamefont {Jin}}, \bibinfo {author}
  {\bibfnamefont {M.}~\bibnamefont {Zhang}},\ and\ \bibinfo {author}
  {\bibfnamefont {J.}~\bibnamefont {Wang}},\ }\bibfield  {title} {\bibinfo
  {title} {Observation of diffusion and drift of the negative trions in
  monolayer {W}{S}\textsubscript{2}},\ }\href
  {https://doi.org/10.1021/acs.nanolett.1c02351} {\bibfield  {journal}
  {\bibinfo  {journal} {Nano Letters}\ }\textbf {\bibinfo {volume} {21}},\
  \bibinfo {pages} {6314} (\bibinfo {year} {2021})}\BibitemShut {NoStop}%
\bibitem [{\citenamefont {Ayari}\ \emph {et~al.}(2020)\citenamefont {Ayari},
  \citenamefont {Jaziri}, \citenamefont {Ferreira},\ and\ \citenamefont
  {Bastard}}]{ayari2020phonon}%
  \BibitemOpen
  \bibfield  {author} {\bibinfo {author} {\bibfnamefont {S.}~\bibnamefont
  {Ayari}}, \bibinfo {author} {\bibfnamefont {S.}~\bibnamefont {Jaziri}},
  \bibinfo {author} {\bibfnamefont {R.}~\bibnamefont {Ferreira}},\ and\
  \bibinfo {author} {\bibfnamefont {G.}~\bibnamefont {Bastard}},\ }\bibfield
  {title} {\bibinfo {title} {Phonon-assisted exciton/trion conversion
  efficiency in transition metal dichalcogenides},\ }\href
  {https://doi.org/10.1103/PhysRevB.102.125410} {\bibfield  {journal} {\bibinfo
   {journal} {Physical Review B}\ }\textbf {\bibinfo {volume} {102}},\ \bibinfo
  {pages} {125410} (\bibinfo {year} {2020})}\BibitemShut {NoStop}%
\bibitem [{\citenamefont {Zipfel}\ \emph {et~al.}(2022)\citenamefont {Zipfel},
  \citenamefont {Wagner}, \citenamefont {Semina}, \citenamefont {Ziegler},
  \citenamefont {Taniguchi}, \citenamefont {Watanabe}, \citenamefont {Glazov},\
  and\ \citenamefont {Chernikov}}]{zipfel2022electron}%
  \BibitemOpen
  \bibfield  {author} {\bibinfo {author} {\bibfnamefont {J.}~\bibnamefont
  {Zipfel}}, \bibinfo {author} {\bibfnamefont {K.}~\bibnamefont {Wagner}},
  \bibinfo {author} {\bibfnamefont {M.~A.}\ \bibnamefont {Semina}}, \bibinfo
  {author} {\bibfnamefont {J.~D.}\ \bibnamefont {Ziegler}}, \bibinfo {author}
  {\bibfnamefont {T.}~\bibnamefont {Taniguchi}}, \bibinfo {author}
  {\bibfnamefont {K.}~\bibnamefont {Watanabe}}, \bibinfo {author}
  {\bibfnamefont {M.~M.}\ \bibnamefont {Glazov}},\ and\ \bibinfo {author}
  {\bibfnamefont {A.}~\bibnamefont {Chernikov}},\ }\bibfield  {title} {\bibinfo
  {title} {Electron recoil effect in electrically tunable
  {M}o{S}e\textsubscript{2} monolayers},\ }\href
  {https://doi.org/10.1103/PhysRevB.105.075311} {\bibfield  {journal} {\bibinfo
   {journal} {Physical Review B}\ }\textbf {\bibinfo {volume} {105}},\ \bibinfo
  {pages} {075311} (\bibinfo {year} {2022})}\BibitemShut {NoStop}%
\bibitem [{\citenamefont {Katsch}\ \emph {et~al.}(2018)\citenamefont {Katsch},
  \citenamefont {Selig}, \citenamefont {Carmele},\ and\ \citenamefont
  {Knorr}}]{katsch2018theory}%
  \BibitemOpen
  \bibfield  {author} {\bibinfo {author} {\bibfnamefont {F.}~\bibnamefont
  {Katsch}}, \bibinfo {author} {\bibfnamefont {M.}~\bibnamefont {Selig}},
  \bibinfo {author} {\bibfnamefont {A.}~\bibnamefont {Carmele}},\ and\ \bibinfo
  {author} {\bibfnamefont {A.}~\bibnamefont {Knorr}},\ }\bibfield  {title}
  {\bibinfo {title} {Theory of exciton--exciton interactions in monolayer
  transition metal dichalcogenides},\ }\href
  {https://doi.org/10.1002/pssb.201800185} {\bibfield  {journal} {\bibinfo
  {journal} {physica status solidi (b)}\ }\textbf {\bibinfo {volume} {255}},\
  \bibinfo {pages} {1800185} (\bibinfo {year} {2018})}\BibitemShut {NoStop}%
\bibitem [{\citenamefont {Ivanov}\ and\ \citenamefont
  {Haug}(1993)}]{ivanov1993self}%
  \BibitemOpen
  \bibfield  {author} {\bibinfo {author} {\bibfnamefont {A.}~\bibnamefont
  {Ivanov}}\ and\ \bibinfo {author} {\bibfnamefont {H.}~\bibnamefont {Haug}},\
  }\bibfield  {title} {\bibinfo {title} {Self-consistent theory of the
  biexciton optical nonlinearity},\ }\href
  {https://doi.org/10.1103/PhysRevB.48.1490} {\bibfield  {journal} {\bibinfo
  {journal} {Physical Review B}\ }\textbf {\bibinfo {volume} {48}},\ \bibinfo
  {pages} {1490} (\bibinfo {year} {1993})}\BibitemShut {NoStop}%
\bibitem [{\citenamefont {Korm{\'a}nyos}\ \emph {et~al.}(2015)\citenamefont
  {Korm{\'a}nyos}, \citenamefont {Burkard}, \citenamefont {Gmitra},
  \citenamefont {Fabian}, \citenamefont {Z{\'o}lyomi}, \citenamefont
  {Drummond},\ and\ \citenamefont {Fal’ko}}]{kormanyos2015k}%
  \BibitemOpen
  \bibfield  {author} {\bibinfo {author} {\bibfnamefont {A.}~\bibnamefont
  {Korm{\'a}nyos}}, \bibinfo {author} {\bibfnamefont {G.}~\bibnamefont
  {Burkard}}, \bibinfo {author} {\bibfnamefont {M.}~\bibnamefont {Gmitra}},
  \bibinfo {author} {\bibfnamefont {J.}~\bibnamefont {Fabian}}, \bibinfo
  {author} {\bibfnamefont {V.}~\bibnamefont {Z{\'o}lyomi}}, \bibinfo {author}
  {\bibfnamefont {N.~D.}\ \bibnamefont {Drummond}},\ and\ \bibinfo {author}
  {\bibfnamefont {V.}~\bibnamefont {Fal’ko}},\ }\bibfield  {title} {\bibinfo
  {title} {k{\textperiodcentered} p theory for two-dimensional transition metal
  dichalcogenide semiconductors},\ }\href
  {https://doi.org/10.1088/2053-1583/2/2/022001} {\bibfield  {journal}
  {\bibinfo  {journal} {2D Materials}\ }\textbf {\bibinfo {volume} {2}},\
  \bibinfo {pages} {022001} (\bibinfo {year} {2015})}\BibitemShut {NoStop}%
\bibitem [{\citenamefont {Yu}\ \emph {et~al.}(2014)\citenamefont {Yu},
  \citenamefont {Liu}, \citenamefont {Gong}, \citenamefont {Xu},\ and\
  \citenamefont {Yao}}]{yu2014dirac}%
  \BibitemOpen
  \bibfield  {author} {\bibinfo {author} {\bibfnamefont {H.}~\bibnamefont
  {Yu}}, \bibinfo {author} {\bibfnamefont {G.-B.}\ \bibnamefont {Liu}},
  \bibinfo {author} {\bibfnamefont {P.}~\bibnamefont {Gong}}, \bibinfo {author}
  {\bibfnamefont {X.}~\bibnamefont {Xu}},\ and\ \bibinfo {author}
  {\bibfnamefont {W.}~\bibnamefont {Yao}},\ }\bibfield  {title} {\bibinfo
  {title} {Dirac cones and dirac saddle points of bright excitons in monolayer
  transition metal dichalcogenides},\ }\href
  {https://doi.org/10.1038/ncomms4876} {\bibfield  {journal} {\bibinfo
  {journal} {Nature Communications}\ }\textbf {\bibinfo {volume} {5}},\
  \bibinfo {pages} {1} (\bibinfo {year} {2014})}\BibitemShut {NoStop}%
\bibitem [{\citenamefont {Van~Tuan}\ \emph {et~al.}(2018)\citenamefont
  {Van~Tuan}, \citenamefont {Yang},\ and\ \citenamefont
  {Dery}}]{van2018coulomb}%
  \BibitemOpen
  \bibfield  {author} {\bibinfo {author} {\bibfnamefont {D.}~\bibnamefont
  {Van~Tuan}}, \bibinfo {author} {\bibfnamefont {M.}~\bibnamefont {Yang}},\
  and\ \bibinfo {author} {\bibfnamefont {H.}~\bibnamefont {Dery}},\ }\bibfield
  {title} {\bibinfo {title} {Coulomb interaction in monolayer transition-metal
  dichalcogenides},\ }\href {https://doi.org/10.1103/PhysRevB.98.125308}
  {\bibfield  {journal} {\bibinfo  {journal} {Physical Review B}\ }\textbf
  {\bibinfo {volume} {98}},\ \bibinfo {pages} {125308} (\bibinfo {year}
  {2018})}\BibitemShut {NoStop}%
\bibitem [{\citenamefont {Hotta}\ \emph {et~al.}(2020)\citenamefont {Hotta},
  \citenamefont {Higuchi}, \citenamefont {Ueda}, \citenamefont {Shinokita},
  \citenamefont {Miyauchi}, \citenamefont {Matsuda}, \citenamefont {Ueno},
  \citenamefont {Taniguchi}, \citenamefont {Watanabe},\ and\ \citenamefont
  {Kitaura}}]{hotta2020exciton}%
  \BibitemOpen
  \bibfield  {author} {\bibinfo {author} {\bibfnamefont {T.}~\bibnamefont
  {Hotta}}, \bibinfo {author} {\bibfnamefont {S.}~\bibnamefont {Higuchi}},
  \bibinfo {author} {\bibfnamefont {A.}~\bibnamefont {Ueda}}, \bibinfo {author}
  {\bibfnamefont {K.}~\bibnamefont {Shinokita}}, \bibinfo {author}
  {\bibfnamefont {Y.}~\bibnamefont {Miyauchi}}, \bibinfo {author}
  {\bibfnamefont {K.}~\bibnamefont {Matsuda}}, \bibinfo {author} {\bibfnamefont
  {K.}~\bibnamefont {Ueno}}, \bibinfo {author} {\bibfnamefont {T.}~\bibnamefont
  {Taniguchi}}, \bibinfo {author} {\bibfnamefont {K.}~\bibnamefont
  {Watanabe}},\ and\ \bibinfo {author} {\bibfnamefont {R.}~\bibnamefont
  {Kitaura}},\ }\bibfield  {title} {\bibinfo {title} {Exciton diffusion in
  h{BN}-encapsulated monolayer {M}o{S}e\textsubscript{2}},\ }\href
  {https://doi.org/10.1103/PhysRevB.102.115424} {\bibfield  {journal} {\bibinfo
   {journal} {Physical Review B}\ }\textbf {\bibinfo {volume} {102}},\ \bibinfo
  {pages} {115424} (\bibinfo {year} {2020})}\BibitemShut {NoStop}%
\bibitem [{\citenamefont {Florian}\ \emph {et~al.}(2018)\citenamefont
  {Florian}, \citenamefont {Hartmann}, \citenamefont {Steinhoff}, \citenamefont
  {Klein}, \citenamefont {Holleitner}, \citenamefont {Finley}, \citenamefont
  {Wehling}, \citenamefont {Kaniber},\ and\ \citenamefont
  {Gies}}]{florian2018dielectric}%
  \BibitemOpen
  \bibfield  {author} {\bibinfo {author} {\bibfnamefont {M.}~\bibnamefont
  {Florian}}, \bibinfo {author} {\bibfnamefont {M.}~\bibnamefont {Hartmann}},
  \bibinfo {author} {\bibfnamefont {A.}~\bibnamefont {Steinhoff}}, \bibinfo
  {author} {\bibfnamefont {J.}~\bibnamefont {Klein}}, \bibinfo {author}
  {\bibfnamefont {A.~W.}\ \bibnamefont {Holleitner}}, \bibinfo {author}
  {\bibfnamefont {J.~J.}\ \bibnamefont {Finley}}, \bibinfo {author}
  {\bibfnamefont {T.~O.}\ \bibnamefont {Wehling}}, \bibinfo {author}
  {\bibfnamefont {M.}~\bibnamefont {Kaniber}},\ and\ \bibinfo {author}
  {\bibfnamefont {C.}~\bibnamefont {Gies}},\ }\bibfield  {title} {\bibinfo
  {title} {The dielectric impact of layer distances on exciton and trion
  binding energies in van der waals heterostructures},\ }\href@noop {}
  {\bibfield  {journal} {\bibinfo  {journal} {Nano Letters}\ }\textbf {\bibinfo
  {volume} {18}},\ \bibinfo {pages} {2725} (\bibinfo {year}
  {2018})}\BibitemShut {NoStop}%
\bibitem [{\citenamefont {Kuhn}\ and\ \citenamefont
  {Rossi}(1992)}]{kuhn1992monte}%
  \BibitemOpen
  \bibfield  {author} {\bibinfo {author} {\bibfnamefont {T.}~\bibnamefont
  {Kuhn}}\ and\ \bibinfo {author} {\bibfnamefont {F.}~\bibnamefont {Rossi}},\
  }\bibfield  {title} {\bibinfo {title} {Monte {C}arlo simulation of ultrafast
  processes in photoexcited semiconductors: Coherent and incoherent dynamics},\
  }\href {https://doi.org/10.1103/PhysRevB.46.7496} {\bibfield  {journal}
  {\bibinfo  {journal} {Physical Review B}\ }\textbf {\bibinfo {volume} {46}},\
  \bibinfo {pages} {7496} (\bibinfo {year} {1992})}\BibitemShut {NoStop}%
\bibitem [{\citenamefont {Brem}\ \emph {et~al.}(2020)\citenamefont {Brem},
  \citenamefont {Ekman}, \citenamefont {Christiansen}, \citenamefont {Katsch},
  \citenamefont {Selig}, \citenamefont {Robert}, \citenamefont {Marie},
  \citenamefont {Urbaszek}, \citenamefont {Knorr},\ and\ \citenamefont
  {Malic}}]{brem20}%
  \BibitemOpen
  \bibfield  {author} {\bibinfo {author} {\bibfnamefont {S.}~\bibnamefont
  {Brem}}, \bibinfo {author} {\bibfnamefont {A.}~\bibnamefont {Ekman}},
  \bibinfo {author} {\bibfnamefont {D.}~\bibnamefont {Christiansen}}, \bibinfo
  {author} {\bibfnamefont {F.}~\bibnamefont {Katsch}}, \bibinfo {author}
  {\bibfnamefont {M.}~\bibnamefont {Selig}}, \bibinfo {author} {\bibfnamefont
  {C.}~\bibnamefont {Robert}}, \bibinfo {author} {\bibfnamefont
  {X.}~\bibnamefont {Marie}}, \bibinfo {author} {\bibfnamefont
  {B.}~\bibnamefont {Urbaszek}}, \bibinfo {author} {\bibfnamefont
  {A.}~\bibnamefont {Knorr}},\ and\ \bibinfo {author} {\bibfnamefont
  {E.}~\bibnamefont {Malic}},\ }\bibfield  {title} {\bibinfo {title}
  {Phonon-assisted photoluminescence from indirect excitons in monolayers of
  transition-metal dichalcogenides},\ }\href
  {https://doi.org/10.1021/acs.nanolett.0c00633} {\bibfield  {journal}
  {\bibinfo  {journal} {Nano letters}\ }\textbf {\bibinfo {volume} {20}},\
  \bibinfo {pages} {2849} (\bibinfo {year} {2020})}\BibitemShut {NoStop}%
\bibitem [{\citenamefont {Jin}\ \emph {et~al.}(2014)\citenamefont {Jin},
  \citenamefont {Li}, \citenamefont {Mullen},\ and\ \citenamefont
  {Kim}}]{jin2014intrinsic}%
  \BibitemOpen
  \bibfield  {author} {\bibinfo {author} {\bibfnamefont {Z.}~\bibnamefont
  {Jin}}, \bibinfo {author} {\bibfnamefont {X.}~\bibnamefont {Li}}, \bibinfo
  {author} {\bibfnamefont {J.~T.}\ \bibnamefont {Mullen}},\ and\ \bibinfo
  {author} {\bibfnamefont {K.~W.}\ \bibnamefont {Kim}},\ }\bibfield  {title}
  {\bibinfo {title} {Intrinsic transport properties of electrons and holes in
  monolayer transition-metal dichalcogenides},\ }\href
  {https://doi.org/10.1103/PhysRevB.90.045422} {\bibfield  {journal} {\bibinfo
  {journal} {Physical Review B}\ }\textbf {\bibinfo {volume} {90}},\ \bibinfo
  {pages} {045422} (\bibinfo {year} {2014})}\BibitemShut {NoStop}%
\bibitem [{\citenamefont {Selig}\ \emph {et~al.}(2016)\citenamefont {Selig},
  \citenamefont {Bergh{\"a}user}, \citenamefont {Raja}, \citenamefont {Nagler},
  \citenamefont {Sch{\"u}ller}, \citenamefont {Heinz}, \citenamefont {Korn},
  \citenamefont {Chernikov}, \citenamefont {Malic},\ and\ \citenamefont
  {Knorr}}]{selig16}%
  \BibitemOpen
  \bibfield  {author} {\bibinfo {author} {\bibfnamefont {M.}~\bibnamefont
  {Selig}}, \bibinfo {author} {\bibfnamefont {G.}~\bibnamefont
  {Bergh{\"a}user}}, \bibinfo {author} {\bibfnamefont {A.}~\bibnamefont
  {Raja}}, \bibinfo {author} {\bibfnamefont {P.}~\bibnamefont {Nagler}},
  \bibinfo {author} {\bibfnamefont {C.}~\bibnamefont {Sch{\"u}ller}}, \bibinfo
  {author} {\bibfnamefont {T.~F.}\ \bibnamefont {Heinz}}, \bibinfo {author}
  {\bibfnamefont {T.}~\bibnamefont {Korn}}, \bibinfo {author} {\bibfnamefont
  {A.}~\bibnamefont {Chernikov}}, \bibinfo {author} {\bibfnamefont
  {E.}~\bibnamefont {Malic}},\ and\ \bibinfo {author} {\bibfnamefont
  {A.}~\bibnamefont {Knorr}},\ }\bibfield  {title} {\bibinfo {title} {Excitonic
  linewidth and coherence lifetime in monolayer transition metal
  dichalcogenides},\ }\href {https://doi.org/10.1038/ncomms13279} {\bibfield
  {journal} {\bibinfo  {journal} {Nature communications}\ }\textbf {\bibinfo
  {volume} {7}},\ \bibinfo {pages} {1} (\bibinfo {year} {2016})}\BibitemShut
  {NoStop}%
\bibitem [{\citenamefont {Brem}\ \emph {et~al.}(2019)\citenamefont {Brem},
  \citenamefont {Zipfel}, \citenamefont {Selig}, \citenamefont {Raja},
  \citenamefont {Waldecker}, \citenamefont {Ziegler}, \citenamefont
  {Taniguchi}, \citenamefont {Watanabe}, \citenamefont {Chernikov},\ and\
  \citenamefont {Malic}}]{brem19}%
  \BibitemOpen
  \bibfield  {author} {\bibinfo {author} {\bibfnamefont {S.}~\bibnamefont
  {Brem}}, \bibinfo {author} {\bibfnamefont {J.}~\bibnamefont {Zipfel}},
  \bibinfo {author} {\bibfnamefont {M.}~\bibnamefont {Selig}}, \bibinfo
  {author} {\bibfnamefont {A.}~\bibnamefont {Raja}}, \bibinfo {author}
  {\bibfnamefont {L.}~\bibnamefont {Waldecker}}, \bibinfo {author}
  {\bibfnamefont {J.~D.}\ \bibnamefont {Ziegler}}, \bibinfo {author}
  {\bibfnamefont {T.}~\bibnamefont {Taniguchi}}, \bibinfo {author}
  {\bibfnamefont {K.}~\bibnamefont {Watanabe}}, \bibinfo {author}
  {\bibfnamefont {A.}~\bibnamefont {Chernikov}},\ and\ \bibinfo {author}
  {\bibfnamefont {E.}~\bibnamefont {Malic}},\ }\bibfield  {title} {\bibinfo
  {title} {Intrinsic lifetime of higher excitonic states in tungsten diselenide
  monolayers},\ }\href {https://doi.org/10.1039/C9NR04211C} {\bibfield
  {journal} {\bibinfo  {journal} {Nanoscale}\ }\textbf {\bibinfo {volume}
  {11}},\ \bibinfo {pages} {12381} (\bibinfo {year} {2019})}\BibitemShut
  {NoStop}%
\bibitem [{\citenamefont {Kaasbjerg}\ \emph {et~al.}(2014)\citenamefont
  {Kaasbjerg}, \citenamefont {Bhargavi},\ and\ \citenamefont
  {Kubakaddi}}]{kaasbjerg2014hot}%
  \BibitemOpen
  \bibfield  {author} {\bibinfo {author} {\bibfnamefont {K.}~\bibnamefont
  {Kaasbjerg}}, \bibinfo {author} {\bibfnamefont {K.}~\bibnamefont
  {Bhargavi}},\ and\ \bibinfo {author} {\bibfnamefont {S.}~\bibnamefont
  {Kubakaddi}},\ }\bibfield  {title} {\bibinfo {title} {Hot-electron cooling by
  acoustic and optical phonons in monolayers of {M}o{S}\textsubscript{2} and
  other transition-metal dichalcogenides},\ }\href
  {https://doi.org/10.1103/PhysRevB.90.165436} {\bibfield  {journal} {\bibinfo
  {journal} {Physical Review B}\ }\textbf {\bibinfo {volume} {90}},\ \bibinfo
  {pages} {165436} (\bibinfo {year} {2014})}\BibitemShut {NoStop}%
\bibitem [{\citenamefont {Venanzi}\ \emph {et~al.}(2021)\citenamefont
  {Venanzi}, \citenamefont {Selig}, \citenamefont {Winnerl}, \citenamefont
  {Pashkin}, \citenamefont {Knorr}, \citenamefont {Helm},\ and\ \citenamefont
  {Schneider}}]{venanzi2021terahertz}%
  \BibitemOpen
  \bibfield  {author} {\bibinfo {author} {\bibfnamefont {T.}~\bibnamefont
  {Venanzi}}, \bibinfo {author} {\bibfnamefont {M.}~\bibnamefont {Selig}},
  \bibinfo {author} {\bibfnamefont {S.}~\bibnamefont {Winnerl}}, \bibinfo
  {author} {\bibfnamefont {A.}~\bibnamefont {Pashkin}}, \bibinfo {author}
  {\bibfnamefont {A.}~\bibnamefont {Knorr}}, \bibinfo {author} {\bibfnamefont
  {M.}~\bibnamefont {Helm}},\ and\ \bibinfo {author} {\bibfnamefont
  {H.}~\bibnamefont {Schneider}},\ }\bibfield  {title} {\bibinfo {title}
  {Terahertz-induced energy transfer from hot carriers to trions in a mose2
  monolayer},\ }\href@noop {} {\bibfield  {journal} {\bibinfo  {journal} {ACS
  Photonics}\ }\textbf {\bibinfo {volume} {8}},\ \bibinfo {pages} {2931}
  (\bibinfo {year} {2021})}\BibitemShut {NoStop}%
\bibitem [{\citenamefont {Hess}\ and\ \citenamefont
  {Kuhn}(1996)}]{hess1996maxwell}%
  \BibitemOpen
  \bibfield  {author} {\bibinfo {author} {\bibfnamefont {O.}~\bibnamefont
  {Hess}}\ and\ \bibinfo {author} {\bibfnamefont {T.}~\bibnamefont {Kuhn}},\
  }\bibfield  {title} {\bibinfo {title} {Maxwell-{B}loch equations for
  spatially inhomogeneous semiconductor lasers. i. theoretical formulation},\
  }\href {https://doi.org/10.1103/PhysRevA.54.3347} {\bibfield  {journal}
  {\bibinfo  {journal} {Physical Review A}\ }\textbf {\bibinfo {volume} {54}},\
  \bibinfo {pages} {3347} (\bibinfo {year} {1996})}\BibitemShut {NoStop}%
\bibitem [{\citenamefont {Rosati}\ \emph {et~al.}(2020)\citenamefont {Rosati},
  \citenamefont {Brem}, \citenamefont {Perea-Caus{\'\i}n}, \citenamefont
  {Schmidt}, \citenamefont {Niehues}, \citenamefont {de~Vasconcellos},
  \citenamefont {Bratschitsch},\ and\ \citenamefont
  {Malic}}]{rosati2020strain}%
  \BibitemOpen
  \bibfield  {author} {\bibinfo {author} {\bibfnamefont {R.}~\bibnamefont
  {Rosati}}, \bibinfo {author} {\bibfnamefont {S.}~\bibnamefont {Brem}},
  \bibinfo {author} {\bibfnamefont {R.}~\bibnamefont {Perea-Caus{\'\i}n}},
  \bibinfo {author} {\bibfnamefont {R.}~\bibnamefont {Schmidt}}, \bibinfo
  {author} {\bibfnamefont {I.}~\bibnamefont {Niehues}}, \bibinfo {author}
  {\bibfnamefont {S.~M.}\ \bibnamefont {de~Vasconcellos}}, \bibinfo {author}
  {\bibfnamefont {R.}~\bibnamefont {Bratschitsch}},\ and\ \bibinfo {author}
  {\bibfnamefont {E.}~\bibnamefont {Malic}},\ }\bibfield  {title} {\bibinfo
  {title} {Strain-dependent exciton diffusion in transition metal
  dichalcogenides},\ }\href {https://doi.org/10.1088/2053-1583/abbd51}
  {\bibfield  {journal} {\bibinfo  {journal} {2D Materials}\ }\textbf {\bibinfo
  {volume} {8}},\ \bibinfo {pages} {015030} (\bibinfo {year}
  {2020})}\BibitemShut {NoStop}%
\bibitem [{\citenamefont {Glazov}(2020{\natexlab{b}})}]{glazov2020quantum}%
  \BibitemOpen
  \bibfield  {author} {\bibinfo {author} {\bibfnamefont {M.}~\bibnamefont
  {Glazov}},\ }\bibfield  {title} {\bibinfo {title} {Quantum interference
  effect on exciton transport in monolayer semiconductors},\ }\href
  {https://doi.org/10.1103/PhysRevLett.124.166802} {\bibfield  {journal}
  {\bibinfo  {journal} {Physical Review Letters}\ }\textbf {\bibinfo {volume}
  {124}},\ \bibinfo {pages} {166802} (\bibinfo {year}
  {2020}{\natexlab{b}})}\BibitemShut {NoStop}%
\bibitem [{\citenamefont {Wagner}\ \emph {et~al.}(2021)\citenamefont {Wagner},
  \citenamefont {Zipfel}, \citenamefont {Rosati}, \citenamefont {Wietek},
  \citenamefont {Ziegler}, \citenamefont {Brem}, \citenamefont
  {Perea-Caus{\'\i}n}, \citenamefont {Taniguchi}, \citenamefont {Watanabe},
  \citenamefont {Glazov} \emph {et~al.}}]{wagner2021nonclassical}%
  \BibitemOpen
  \bibfield  {author} {\bibinfo {author} {\bibfnamefont {K.}~\bibnamefont
  {Wagner}}, \bibinfo {author} {\bibfnamefont {J.}~\bibnamefont {Zipfel}},
  \bibinfo {author} {\bibfnamefont {R.}~\bibnamefont {Rosati}}, \bibinfo
  {author} {\bibfnamefont {E.}~\bibnamefont {Wietek}}, \bibinfo {author}
  {\bibfnamefont {J.~D.}\ \bibnamefont {Ziegler}}, \bibinfo {author}
  {\bibfnamefont {S.}~\bibnamefont {Brem}}, \bibinfo {author} {\bibfnamefont
  {R.}~\bibnamefont {Perea-Caus{\'\i}n}}, \bibinfo {author} {\bibfnamefont
  {T.}~\bibnamefont {Taniguchi}}, \bibinfo {author} {\bibfnamefont
  {K.}~\bibnamefont {Watanabe}}, \bibinfo {author} {\bibfnamefont {M.~M.}\
  \bibnamefont {Glazov}}, \emph {et~al.},\ }\bibfield  {title} {\bibinfo
  {title} {Nonclassical exciton diffusion in monolayer
  {W}{S}e\textsubscript{2}},\ }\href
  {https://doi.org/10.1103/PhysRevLett.127.076801} {\bibfield  {journal}
  {\bibinfo  {journal} {Physical Review Letters}\ }\textbf {\bibinfo {volume}
  {127}},\ \bibinfo {pages} {076801} (\bibinfo {year} {2021})}\BibitemShut
  {NoStop}%
\bibitem [{\citenamefont {Thompson}\ \emph {et~al.}(2022)\citenamefont
  {Thompson}, \citenamefont {Brem}, \citenamefont {Verjans}, \citenamefont
  {Schmidt}, \citenamefont {de~Vasconcellos}, \citenamefont {Bratschitsch},\
  and\ \citenamefont {Malic}}]{thompson2022anisotropic}%
  \BibitemOpen
  \bibfield  {author} {\bibinfo {author} {\bibfnamefont {J.~J.}\ \bibnamefont
  {Thompson}}, \bibinfo {author} {\bibfnamefont {S.}~\bibnamefont {Brem}},
  \bibinfo {author} {\bibfnamefont {M.}~\bibnamefont {Verjans}}, \bibinfo
  {author} {\bibfnamefont {R.}~\bibnamefont {Schmidt}}, \bibinfo {author}
  {\bibfnamefont {S.~M.}\ \bibnamefont {de~Vasconcellos}}, \bibinfo {author}
  {\bibfnamefont {R.}~\bibnamefont {Bratschitsch}},\ and\ \bibinfo {author}
  {\bibfnamefont {E.}~\bibnamefont {Malic}},\ }\bibfield  {title} {\bibinfo
  {title} {Anisotropic exciton diffusion in atomically-thin semiconductors},\
  }\href@noop {} {\bibfield  {journal} {\bibinfo  {journal} {2D Materials}\
  }\textbf {\bibinfo {volume} {9}},\ \bibinfo {pages} {025008} (\bibinfo {year}
  {2022})}\BibitemShut {NoStop}%
\bibitem [{\citenamefont {Rosati}\ \emph {et~al.}(2021)\citenamefont {Rosati},
  \citenamefont {Schmidt}, \citenamefont {Brem}, \citenamefont
  {Perea-Caus{\'\i}n}, \citenamefont {Niehues}, \citenamefont {Kern},
  \citenamefont {Preu{\ss}}, \citenamefont {Schneider}, \citenamefont
  {Michaelis~de Vasconcellos}, \citenamefont {Bratschitsch} \emph
  {et~al.}}]{rosati2021dark}%
  \BibitemOpen
  \bibfield  {author} {\bibinfo {author} {\bibfnamefont {R.}~\bibnamefont
  {Rosati}}, \bibinfo {author} {\bibfnamefont {R.}~\bibnamefont {Schmidt}},
  \bibinfo {author} {\bibfnamefont {S.}~\bibnamefont {Brem}}, \bibinfo {author}
  {\bibfnamefont {R.}~\bibnamefont {Perea-Caus{\'\i}n}}, \bibinfo {author}
  {\bibfnamefont {I.}~\bibnamefont {Niehues}}, \bibinfo {author} {\bibfnamefont
  {J.}~\bibnamefont {Kern}}, \bibinfo {author} {\bibfnamefont {J.~A.}\
  \bibnamefont {Preu{\ss}}}, \bibinfo {author} {\bibfnamefont {R.}~\bibnamefont
  {Schneider}}, \bibinfo {author} {\bibfnamefont {S.}~\bibnamefont
  {Michaelis~de Vasconcellos}}, \bibinfo {author} {\bibfnamefont
  {R.}~\bibnamefont {Bratschitsch}}, \emph {et~al.},\ }\bibfield  {title}
  {\bibinfo {title} {Dark exciton anti-funneling in atomically thin
  semiconductors},\ }\href {https://doi.org/10.1038/s41467-021-27425-y}
  {\bibfield  {journal} {\bibinfo  {journal} {Nature Communications}\ }\textbf
  {\bibinfo {volume} {12}},\ \bibinfo {pages} {1} (\bibinfo {year}
  {2021})}\BibitemShut {NoStop}%
\bibitem [{\citenamefont {Unuchek}\ \emph {et~al.}(2018)\citenamefont
  {Unuchek}, \citenamefont {Ciarrocchi}, \citenamefont {Avsar}, \citenamefont
  {Watanabe}, \citenamefont {Taniguchi},\ and\ \citenamefont
  {Kis}}]{unuchek2018room}%
  \BibitemOpen
  \bibfield  {author} {\bibinfo {author} {\bibfnamefont {D.}~\bibnamefont
  {Unuchek}}, \bibinfo {author} {\bibfnamefont {A.}~\bibnamefont {Ciarrocchi}},
  \bibinfo {author} {\bibfnamefont {A.}~\bibnamefont {Avsar}}, \bibinfo
  {author} {\bibfnamefont {K.}~\bibnamefont {Watanabe}}, \bibinfo {author}
  {\bibfnamefont {T.}~\bibnamefont {Taniguchi}},\ and\ \bibinfo {author}
  {\bibfnamefont {A.}~\bibnamefont {Kis}},\ }\bibfield  {title} {\bibinfo
  {title} {Room-temperature electrical control of exciton flux in a van der
  {W}aals heterostructure},\ }\href {https://doi.org/10.1038/s41586-018-0357-y}
  {\bibfield  {journal} {\bibinfo  {journal} {Nature}\ }\textbf {\bibinfo
  {volume} {560}},\ \bibinfo {pages} {340} (\bibinfo {year}
  {2018})}\BibitemShut {NoStop}%
\bibitem [{\citenamefont {Lui}\ \emph {et~al.}(2014)\citenamefont {Lui},
  \citenamefont {Frenzel}, \citenamefont {Pilon}, \citenamefont {Lee},
  \citenamefont {Ling}, \citenamefont {Akselrod}, \citenamefont {Kong},\ and\
  \citenamefont {Gedik}}]{lui2014trion}%
  \BibitemOpen
  \bibfield  {author} {\bibinfo {author} {\bibfnamefont {C.}~\bibnamefont
  {Lui}}, \bibinfo {author} {\bibfnamefont {A.}~\bibnamefont {Frenzel}},
  \bibinfo {author} {\bibfnamefont {D.}~\bibnamefont {Pilon}}, \bibinfo
  {author} {\bibfnamefont {Y.-H.}\ \bibnamefont {Lee}}, \bibinfo {author}
  {\bibfnamefont {X.}~\bibnamefont {Ling}}, \bibinfo {author} {\bibfnamefont
  {G.}~\bibnamefont {Akselrod}}, \bibinfo {author} {\bibfnamefont
  {J.}~\bibnamefont {Kong}},\ and\ \bibinfo {author} {\bibfnamefont
  {N.}~\bibnamefont {Gedik}},\ }\bibfield  {title} {\bibinfo {title}
  {Trion-induced negative photoconductivity in monolayer
  {M}o{S}\textsubscript{2}},\ }\href
  {https://doi.org/10.1103/PhysRevLett.113.166801} {\bibfield  {journal}
  {\bibinfo  {journal} {Physical Review Letters}\ }\textbf {\bibinfo {volume}
  {113}},\ \bibinfo {pages} {166801} (\bibinfo {year} {2014})}\BibitemShut
  {NoStop}%
\bibitem [{\citenamefont {Helmrich}\ \emph {et~al.}(2021)\citenamefont
  {Helmrich}, \citenamefont {Achtstein}, \citenamefont {Ahmad}, \citenamefont
  {Kunz}, \citenamefont {Herzog}, \citenamefont {Sch{\"o}ps}, \citenamefont
  {Woggon},\ and\ \citenamefont {Owschimikow}}]{helmrich2021high}%
  \BibitemOpen
  \bibfield  {author} {\bibinfo {author} {\bibfnamefont {S.}~\bibnamefont
  {Helmrich}}, \bibinfo {author} {\bibfnamefont {A.~W.}\ \bibnamefont
  {Achtstein}}, \bibinfo {author} {\bibfnamefont {H.}~\bibnamefont {Ahmad}},
  \bibinfo {author} {\bibfnamefont {M.}~\bibnamefont {Kunz}}, \bibinfo {author}
  {\bibfnamefont {B.}~\bibnamefont {Herzog}}, \bibinfo {author} {\bibfnamefont
  {O.}~\bibnamefont {Sch{\"o}ps}}, \bibinfo {author} {\bibfnamefont
  {U.}~\bibnamefont {Woggon}},\ and\ \bibinfo {author} {\bibfnamefont
  {N.}~\bibnamefont {Owschimikow}},\ }\bibfield  {title} {\bibinfo {title}
  {High phonon-limited mobility of charged and neutral excitons in mono-and
  bilayer {M}o{T}e\textsubscript{2}},\ }\href
  {https://ui.adsabs.harvard.edu/link_gateway/2021TDM.....8b5019H/doi:10.1088/2053-1583/abd827}
  {\bibfield  {journal} {\bibinfo  {journal} {2D Materials}\ }\textbf {\bibinfo
  {volume} {8}},\ \bibinfo {pages} {025019} (\bibinfo {year}
  {2021})}\BibitemShut {NoStop}%
\bibitem [{\citenamefont {Raja}\ \emph {et~al.}(2019)\citenamefont {Raja},
  \citenamefont {Waldecker}, \citenamefont {Zipfel}, \citenamefont {Cho},
  \citenamefont {Brem}, \citenamefont {Ziegler}, \citenamefont {Kulig},
  \citenamefont {Taniguchi}, \citenamefont {Watanabe}, \citenamefont {Malic}
  \emph {et~al.}}]{raja2019dielectric}%
  \BibitemOpen
  \bibfield  {author} {\bibinfo {author} {\bibfnamefont {A.}~\bibnamefont
  {Raja}}, \bibinfo {author} {\bibfnamefont {L.}~\bibnamefont {Waldecker}},
  \bibinfo {author} {\bibfnamefont {J.}~\bibnamefont {Zipfel}}, \bibinfo
  {author} {\bibfnamefont {Y.}~\bibnamefont {Cho}}, \bibinfo {author}
  {\bibfnamefont {S.}~\bibnamefont {Brem}}, \bibinfo {author} {\bibfnamefont
  {J.~D.}\ \bibnamefont {Ziegler}}, \bibinfo {author} {\bibfnamefont
  {M.}~\bibnamefont {Kulig}}, \bibinfo {author} {\bibfnamefont
  {T.}~\bibnamefont {Taniguchi}}, \bibinfo {author} {\bibfnamefont
  {K.}~\bibnamefont {Watanabe}}, \bibinfo {author} {\bibfnamefont
  {E.}~\bibnamefont {Malic}}, \emph {et~al.},\ }\bibfield  {title} {\bibinfo
  {title} {Dielectric disorder in two-dimensional materials},\ }\href
  {https://doi.org/10.1038/s41565-019-0520-0} {\bibfield  {journal} {\bibinfo
  {journal} {Nature Nanotechnology}\ }\textbf {\bibinfo {volume} {14}},\
  \bibinfo {pages} {832} (\bibinfo {year} {2019})}\BibitemShut {NoStop}%
\bibitem [{\citenamefont {Zipfel}\ \emph {et~al.}(2020)\citenamefont {Zipfel},
  \citenamefont {Kulig}, \citenamefont {Perea-Caus{\'\i}n}, \citenamefont
  {Brem}, \citenamefont {Ziegler}, \citenamefont {Rosati}, \citenamefont
  {Taniguchi}, \citenamefont {Watanabe}, \citenamefont {Glazov}, \citenamefont
  {Malic} \emph {et~al.}}]{zipfel2020exciton}%
  \BibitemOpen
  \bibfield  {author} {\bibinfo {author} {\bibfnamefont {J.}~\bibnamefont
  {Zipfel}}, \bibinfo {author} {\bibfnamefont {M.}~\bibnamefont {Kulig}},
  \bibinfo {author} {\bibfnamefont {R.}~\bibnamefont {Perea-Caus{\'\i}n}},
  \bibinfo {author} {\bibfnamefont {S.}~\bibnamefont {Brem}}, \bibinfo {author}
  {\bibfnamefont {J.~D.}\ \bibnamefont {Ziegler}}, \bibinfo {author}
  {\bibfnamefont {R.}~\bibnamefont {Rosati}}, \bibinfo {author} {\bibfnamefont
  {T.}~\bibnamefont {Taniguchi}}, \bibinfo {author} {\bibfnamefont
  {K.}~\bibnamefont {Watanabe}}, \bibinfo {author} {\bibfnamefont {M.~M.}\
  \bibnamefont {Glazov}}, \bibinfo {author} {\bibfnamefont {E.}~\bibnamefont
  {Malic}}, \emph {et~al.},\ }\bibfield  {title} {\bibinfo {title} {Exciton
  diffusion in monolayer semiconductors with suppressed disorder},\ }\href
  {https://doi.org/10.1103/PhysRevB.101.115430} {\bibfield  {journal} {\bibinfo
   {journal} {Physical Review B}\ }\textbf {\bibinfo {volume} {101}},\ \bibinfo
  {pages} {115430} (\bibinfo {year} {2020})}\BibitemShut {NoStop}%
\bibitem [{\citenamefont {Glazov}(2019)}]{glazov2019phonon}%
  \BibitemOpen
  \bibfield  {author} {\bibinfo {author} {\bibfnamefont {M.}~\bibnamefont
  {Glazov}},\ }\bibfield  {title} {\bibinfo {title} {Phonon wind and drag of
  excitons in monolayer semiconductors},\ }\href@noop {} {\bibfield  {journal}
  {\bibinfo  {journal} {Physical Review B}\ }\textbf {\bibinfo {volume}
  {100}},\ \bibinfo {pages} {045426} (\bibinfo {year} {2019})}\BibitemShut
  {NoStop}%
\bibitem [{\citenamefont {Perea-Causin}\ \emph {et~al.}(2019)\citenamefont
  {Perea-Causin}, \citenamefont {Brem}, \citenamefont {Rosati}, \citenamefont
  {Jago}, \citenamefont {Kulig}, \citenamefont {Ziegler}, \citenamefont
  {Zipfel}, \citenamefont {Chernikov},\ and\ \citenamefont
  {Malic}}]{perea2019exciton}%
  \BibitemOpen
  \bibfield  {author} {\bibinfo {author} {\bibfnamefont {R.}~\bibnamefont
  {Perea-Causin}}, \bibinfo {author} {\bibfnamefont {S.}~\bibnamefont {Brem}},
  \bibinfo {author} {\bibfnamefont {R.}~\bibnamefont {Rosati}}, \bibinfo
  {author} {\bibfnamefont {R.}~\bibnamefont {Jago}}, \bibinfo {author}
  {\bibfnamefont {M.}~\bibnamefont {Kulig}}, \bibinfo {author} {\bibfnamefont
  {J.~D.}\ \bibnamefont {Ziegler}}, \bibinfo {author} {\bibfnamefont
  {J.}~\bibnamefont {Zipfel}}, \bibinfo {author} {\bibfnamefont
  {A.}~\bibnamefont {Chernikov}},\ and\ \bibinfo {author} {\bibfnamefont
  {E.}~\bibnamefont {Malic}},\ }\bibfield  {title} {\bibinfo {title} {Exciton
  propagation and halo formation in two-dimensional materials},\ }\href@noop {}
  {\bibfield  {journal} {\bibinfo  {journal} {Nano letters}\ }\textbf {\bibinfo
  {volume} {19}},\ \bibinfo {pages} {7317} (\bibinfo {year}
  {2019})}\BibitemShut {NoStop}%
\bibitem [{\citenamefont {Kira}\ and\ \citenamefont
  {Koch}(2006)}]{kira2006many}%
  \BibitemOpen
  \bibfield  {author} {\bibinfo {author} {\bibfnamefont {M.}~\bibnamefont
  {Kira}}\ and\ \bibinfo {author} {\bibfnamefont {S.~W.}\ \bibnamefont
  {Koch}},\ }\bibfield  {title} {\bibinfo {title} {Many-body correlations and
  excitonic effects in semiconductor spectroscopy},\ }\href
  {https://doi.org/10.1016/j.pquantelec.2006.12.002} {\bibfield  {journal}
  {\bibinfo  {journal} {Progress in Quantum Electronics}\ }\textbf {\bibinfo
  {volume} {30}},\ \bibinfo {pages} {155} (\bibinfo {year} {2006})}\BibitemShut
  {NoStop}%
\bibitem [{\citenamefont {Bergh{\"a}user}\ \emph {et~al.}(2018)\citenamefont
  {Bergh{\"a}user}, \citenamefont {Steinleitner}, \citenamefont {Merkl},
  \citenamefont {Huber}, \citenamefont {Knorr},\ and\ \citenamefont
  {Malic}}]{berghauser2018mapping}%
  \BibitemOpen
  \bibfield  {author} {\bibinfo {author} {\bibfnamefont {G.}~\bibnamefont
  {Bergh{\"a}user}}, \bibinfo {author} {\bibfnamefont {P.}~\bibnamefont
  {Steinleitner}}, \bibinfo {author} {\bibfnamefont {P.}~\bibnamefont {Merkl}},
  \bibinfo {author} {\bibfnamefont {R.}~\bibnamefont {Huber}}, \bibinfo
  {author} {\bibfnamefont {A.}~\bibnamefont {Knorr}},\ and\ \bibinfo {author}
  {\bibfnamefont {E.}~\bibnamefont {Malic}},\ }\bibfield  {title} {\bibinfo
  {title} {Mapping of the dark exciton landscape in transition metal
  dichalcogenides},\ }\href {https://doi.org/10.1103/PhysRevB.98.020301}
  {\bibfield  {journal} {\bibinfo  {journal} {Physical Review B}\ }\textbf
  {\bibinfo {volume} {98}},\ \bibinfo {pages} {020301} (\bibinfo {year}
  {2018})}\BibitemShut {NoStop}%
\bibitem [{\citenamefont {Peelaers}\ and\ \citenamefont {Van~de
  Walle}(2012)}]{peelaers2012effects}%
  \BibitemOpen
  \bibfield  {author} {\bibinfo {author} {\bibfnamefont {H.}~\bibnamefont
  {Peelaers}}\ and\ \bibinfo {author} {\bibfnamefont {C.~G.}\ \bibnamefont
  {Van~de Walle}},\ }\bibfield  {title} {\bibinfo {title} {Effects of strain on
  band structure and effective masses in {M}o{S}\textsubscript{2}},\ }\href
  {http://dx.doi.org/10.1103/PhysRevB.86.241401} {\bibfield  {journal}
  {\bibinfo  {journal} {Physical Review B}\ }\textbf {\bibinfo {volume} {86}},\
  \bibinfo {pages} {241401} (\bibinfo {year} {2012})}\BibitemShut {NoStop}%
\bibitem [{\citenamefont {Shree}\ \emph {et~al.}(2018)\citenamefont {Shree},
  \citenamefont {Semina}, \citenamefont {Robert}, \citenamefont {Han},
  \citenamefont {Amand}, \citenamefont {Balocchi}, \citenamefont {Manca},
  \citenamefont {Courtade}, \citenamefont {Marie}, \citenamefont {Taniguchi}
  \emph {et~al.}}]{shree2018observation}%
  \BibitemOpen
  \bibfield  {author} {\bibinfo {author} {\bibfnamefont {S.}~\bibnamefont
  {Shree}}, \bibinfo {author} {\bibfnamefont {M.}~\bibnamefont {Semina}},
  \bibinfo {author} {\bibfnamefont {C.}~\bibnamefont {Robert}}, \bibinfo
  {author} {\bibfnamefont {B.}~\bibnamefont {Han}}, \bibinfo {author}
  {\bibfnamefont {T.}~\bibnamefont {Amand}}, \bibinfo {author} {\bibfnamefont
  {A.}~\bibnamefont {Balocchi}}, \bibinfo {author} {\bibfnamefont
  {M.}~\bibnamefont {Manca}}, \bibinfo {author} {\bibfnamefont
  {E.}~\bibnamefont {Courtade}}, \bibinfo {author} {\bibfnamefont
  {X.}~\bibnamefont {Marie}}, \bibinfo {author} {\bibfnamefont
  {T.}~\bibnamefont {Taniguchi}}, \emph {et~al.},\ }\bibfield  {title}
  {\bibinfo {title} {Observation of exciton-phonon coupling in
  {M}o{S}e\textsubscript{2} monolayers},\ }\href
  {https://doi.org/10.1103/PhysRevB.98.035302} {\bibfield  {journal} {\bibinfo
  {journal} {Physical Review B}\ }\textbf {\bibinfo {volume} {98}},\ \bibinfo
  {pages} {035302} (\bibinfo {year} {2018})}\BibitemShut {NoStop}%
\bibitem [{\citenamefont {Geick}\ \emph {et~al.}(1966)\citenamefont {Geick},
  \citenamefont {Perry},\ and\ \citenamefont {Rupprecht}}]{geick1966normal}%
  \BibitemOpen
  \bibfield  {author} {\bibinfo {author} {\bibfnamefont {R.}~\bibnamefont
  {Geick}}, \bibinfo {author} {\bibfnamefont {C.}~\bibnamefont {Perry}},\ and\
  \bibinfo {author} {\bibfnamefont {G.}~\bibnamefont {Rupprecht}},\ }\bibfield
  {title} {\bibinfo {title} {Normal modes in hexagonal boron nitride},\ }\href
  {https://doi.org/10.1103/PhysRev.146.543} {\bibfield  {journal} {\bibinfo
  {journal} {Physical Review}\ }\textbf {\bibinfo {volume} {146}},\ \bibinfo
  {pages} {543} (\bibinfo {year} {1966})}\BibitemShut {NoStop}%
\end{thebibliography}

%

\end{document}